\documentclass[twocolumn]{aastex61}

\usepackage{natbib}
\usepackage{amsmath}
\usepackage{epstopdf}
\usepackage{hyperref}

\usepackage{color}
\usepackage{soul}

\newcommand{\vect}[1]{\boldsymbol{#1}}

\begin{document}
\title{GMC Collisions As Triggers of Star Formation. IV.\\The Role of Ambipolar Diffusion}
\author{Duncan Christie}
\affiliation{Department of Astronomy, University of Florida, Gainesville, FL 32611, USA}

\author{Benjamin Wu}
\affiliation{National Astronomical Observatory of Japan, Mitaka, Tokyo 181-8588, Japan}

\author{Jonathan C. Tan}
\affiliation{Department of Astronomy, University of Florida, Gainesville, FL 32611, USA}
\affiliation{Department of Physics, University of Florida, Gainesville, FL 32611, USA}

\begin{abstract}
We investigate the role of ambipolar diffusion (AD) in collisions between
magnetized giant molecular clouds (GMCs), which may be an important
mechanism for triggering star cluster formation.
Three dimensional simulations of GMC collisions are performed using a
version of the \texttt{Enzo} magnetohydrodynamics code that has been
extended to include AD. The resistivities are calculated using the
31-species chemical model of Wu et al. (2015). We find that in the
weak-field, $10\:{\rm \mu G}$ case, AD has only a modest effect on the
dynamical evolution during the collision. However, for the
stronger-field, $30\:{\rm \mu G}$ case involving near-critical clouds,
AD results in formation of dense cores in regions where collapse is
otherwise inhibited. The overall efficiency of formation of cores with
$n_{\rm H}\geq10^{6}\:{\rm cm}^{-3}$ in these simulations is increases
from about 0.2\% to 2\% once AD is included, comparable to observed
values in star-forming GMCs. The gas around these cores typically has
relatively slow infall at speeds that are a modest fraction of the
free-fall speed.
\end{abstract}

\section{Introduction}\label{S:intro}

Collisions between giant molecular clouds (GMCs) have been proposed as
a mechanism to initiate the conditions necessary for gas to form
massive stars and star clusters (e.g.,
\citealt{1986ApJ...310L..77S,2000ApJ...536..173T}). These conditions
are expected to involve there being a large mass of dense gas in
gravitationally unstable clumps and cores, e.g., as potentially traced
by Infrared Dark Clouds (IRDCs) (see, e.g., \citealt{2014prpl.conf..149T}). 
Magnetic fields could play an important role in this process,
especially if they inhibit collapse and regulate the star formation
rate (SFR) in GMCs during the normal circumstances when they are not
colliding (e.g.,
\citealt{1989ApJ...345..782M,2014prpl.conf..101L}). Average SFRs in
GMCs are known to be very inefficient per local free-fall time
\citep{1974ApJ...192L.149Z,2007ApJ...654..304K}, but to also show
large variation \citep{2016ApJ...833..229L}, which could be caused by
triggering by cloud collisions.

During the process of collapse from low density gas conditions to
dense cores and protostars, the mass-to-magnetic flux ratio increases
by many orders of magnitude and a threshold must be crossed of gas
initially being magnetically subcritical (i.e., $B$-fields are strong
enough to resist collapse) to being magnetically super-critical
($B$-fields are not strong enough to stop collapse). The imperfect
coupling between the gas and the magnetic field leads to ambipolar
diffusion (or ion-neutral drift) \citep{1956MNRAS.116..503M}, which is
one way to allow for collapse to be accelerated, especially in
situations where strong field gradients are created such as a
cloud-cloud collision.


Magnetic field strengths within the diffuse interstellar medium (ISM)
have been measured via the Zeeman effect with values of $6\pm 1.8\,
{\rm \mu G}$ \citep{2005ApJ...624..773H} over a wide range of observed
column densities.  The magnetic field strength remains roughly
independent of gas density until a threshold density, $n_{\rm H}\sim
300\,\,{\rm cm}^{-3}$, is reached, after which it begins to increase
with gas density as $B \propto \rho^\kappa$ with $\kappa\simeq 0.65$
 \citep{1999ApJ...520..706C, 2008A&A...487..247F,2010ApJ...725..466C}. This increase can be
interpreted as being due to flux freezing in the collapse of spherical,
weakly-magnetized clouds \citep{1966MNRAS.133..265M}. Such a dependence of $B$-field
strength with density has also been seen in the simulations of
self-gravitating turbulence of, e.g., \citet{2015MNRAS.452.2500L} and
\citet{2017ApJ...838...40M}.
In more strongly magnetized clouds, i.e., with Alfv\'en Mach numbers
$\lesssim 1$, a shallower relation, with $\kappa\simeq0.5$ is expected \citep{2015MNRAS.451.4384T,2017ApJ...838...40M}
and \citet{2015MNRAS.451.4384T} have
also argued that such a value of $\kappa$ is a better match to the
observational data.

Concerning the collapse of GMCs, clumps and massive cores, since
typical temperatures are $\sim 15\,{\rm K}$, thermal pressure
alone is incapable of providing significant support against
gravity. Both magnetic fields and turbulence are in principle capable
of supporting the clouds, if sufficiently strong.  The degree of
support provided by the magnetic field is measured by the mass-to-flux
ratio $\mu$,
\begin{equation}
\mu = \frac{M/\Phi}{1/\sqrt{63G}},
\label{Eqn:MTF}
\end{equation}
where $M$ is the cloud mass and $\Phi$ is the magnetic flux through
the cloud. Here $\mu$ has been normalized to the critical value, so
that $\mu < 1$ is the condition for stability against collapse in the
limit of ideal MHD \citep{1976ApJ...210..326M}. In the case of
initially subcritical clouds, ambipolar diffusion allows gas to be
redistributed across flux tubes, creating structures that are locally
supercritical and unstable.  For sufficiently supercritical
clouds, ambipolar diffusion is not requisite for
collapse\footnote{The mass-to-flux ratio as a gauge of stability
    ignores thermal pressure which can shift the value of $\mu$ at
    which collapse occurs.  Linear analysis and the magneto-Jeans
    instability better captures this (see, e.g.,
    \citealt{2011MNRAS.415.1751M}); however, for simplicity we make
    the approximation that $\mu = 1$ represents the instability
    threshold.} to occur, especially if large magnetic field
and/or velocity gradients are created in the cloud. One mechanism that
may induce such gradients is GMC collisions.


Observationally, inferring the mass-to-flux ratio in interstellar
clouds is difficult due to projection effects associated with
measuring the magnetic field and the column density along the field
line.  \citet{2009ApJ...692..844C} attempted to measure the change in
the projected mass-to-flux ratio between cores and their surrounding
envelopes and concluded that the mass-to-flux ratio decreases from the
envelope to core.  An increase in the mass-to-flux ratio from the
envelope to the core is expected to be a sign of ambipolar diffusion
acting in subcritical clouds, a result not favored by the
\citet{2009ApJ...692..844C} analysis.  However,
\citet{2009MNRAS.400L..15M,2010MNRAS.409..801M} have called into
question the validity of the conclusions on statistical grounds (but,
see also \citealt{2015MNRAS.452.2500L}).



\citet{2008ApJ...677.1151L} proposed that the action of ambipolar
diffusion could be inferred through the differing velocity dispersions
observed for spatially coincident ions and neutrals.  At scales above
the ambipolar diffusion dissipation scale, the ions and neutrals are
expected to remain well-coupled, resulting in near-identical power
spectra for the two species.  Below the dissipation scale, however,
ion modes are quickly damped while the neutrals experience negligible
drag due to the ions.  While it may not be feasible to observe the
dissipation scale directly, the velocity dispersion retains
information about all scales smaller than the beam size.  The
technique has been applied multiple times with the resulting
dissipation scales being on the order of $10^{-3}$ to $10^{-2}\,{\rm
  pc}$ \citep{2010ApJ...718..905L,2010ApJ...720..603H,2014MNRAS.438..663H}.

In this paper we investigate the role ambipolar diffusion plays during
the process of GMC-GMC collisions. \citet[hereafter Paper
  I]{2015ApJ...811...56W} and \citet[hereafter Paper
  II]{2017ApJ...835..137W} investigated how such collisions can
trigger dense clump formation, including magnetic fields treated
under the assumption of ideal magneto-hydrodynamics (MHD) and a model
of the multiphase interstellar medium. In \citet[hereafter Paper
  III]{2017ApJ...841...88W}, star formation was explicitly included in
the simulations with various sub-grid models investigated, including
examples in which a mass-to-flux threshold criterion needs to be
reached in order for gas in a cell to be able to form stars.


This work expands upon Papers I and II through the inclusion of
ambipolar diffusion, modifications to the heating and cooling model,
and the expansion to larger effective resolutions.  Section \ref{Sect:Model}
describes the numerical model, including the initial conditions
(Section \ref{Sect:InitCond}), the treatment of heating and cooling
(Section \ref{Sect:HeatCool}), the inclusion of ambipolar diffusion
(Section \ref{Sect:AD}), and the calculation of resistivities
(Section \ref{Sect:Resist}).  The results are presented in
Section \ref{Sect:Results} and the conclusions in Section \ref{Sect:Conclusions}.
Tests of the ambipolar diffusion module are described in Appendix
\ref{Append:AD}.

\section{Model of Colliding and Non-Colliding GMCs}
\label{Sect:Model}

The model presented here is very similar to that presented in Paper
II, with the differences confined to updates to the heating and
cooling model and the inclusion of ambipolar diffusion.

\subsection{Initial Conditions}
\label{Sect:InitCond}

The simulation volume is a $L=128\,{\rm pc}$ sided cube filled with an
ambient medium with $n_{\rm H,0} = 10\,{\rm cm^{-3}}$.  Two identical,
spherical GMCs with radii $R_{\rm GMC} = 20\,{\rm pc}$ and $n_{\rm
  H,0} = 100\,{\rm cm^{-3}}$ are embedded within the ambient medium
and offset with an impact parameter $b=0.5 R_{\rm GMC}$.  In addition
to the hydrogen component, the gas contains heavier species in
abundances representative of the local interstellar medium.

Within each model cloud, a turbulent velocity field is initialized
with a spectrum $v_{\rm k}^2\propto k^{-4}$, where $k$ is the
wavenumber.  For gas with $T=15\,{\rm K}$, the virial scale Mach
number $\mathcal{M}_s = \sigma/c_{\rm s}=23$.  The $k$-modes are
chosen to be in the range $2 < k/(\pi/L)< 20$.  In addition to the
turbulent velocity spectrum, the clouds are imparted with a relative
velocity $v_{\rm rel}$ with the fiducial value being $v_{\rm rel} =
10\,{\rm km\, s^{-1}}$, although we also investigate the stationary
(non-colliding) case $v_{\rm rel} = 0\,{\rm km\, s^{-1}}$.  
In the simulation frame of reference, half of the gas is given a velocity of
$+v_{\rm rel}/2$, while the other half given a velocity of
$-v_{\rm rel}/2$.

A uniform magnetic field with field strength $B=10\,{\rm \mu G}$ or
$30\,{\rm \mu G}$ is initialized with a direction at an angle
$\theta=60^\circ$ relative to the collision axis of the clouds.
Ignoring the contribution from the ambient medium and taking the flux
through the cloud to be $\Phi = \pi R_{\rm GMC}^2 B_0$, the initial
mass-to-flux ratios are $\mu_0 = 3.8$ for the $10\,{\rm \mu G}$ runs
and $\mu_0 = 1.3$ for the $30\,{\rm \mu G}$ runs, both supercritical
and globally unstable against collapse. Due to their large extents,
each cloud may fragment along the field lines, creating objects with
smaller, potentially subcritical, mass-to-flux ratios.

Due to the spherical geometry adopted for the model GMCs, the central
flux tube is loaded with more mass resulting in a central mass-to-flux
ratio, $\mu_{\rm center}$, of
\begin{equation}
\mu_{\rm center} \sim \frac{N/B_0}{1/\sqrt{63G}} = \frac{2R_{\rm GMC}\rho_0/B_0}{1/\sqrt{63G}} = \left\{ \begin{matrix}
5.7, & B=10\,{\rm \mu G}\\
1.9, & B=30\,{\rm \mu G} 
\end{matrix} \right.\,\,.
\end{equation}
The larger values of $\mu$ associated with the central flux tubes
through the cloud could potentially result in preferential collapse in
these regions compared to near the cloud boundaries.

To investigate the role of ambipolar diffusion, identical initial
conditions are run with ambipolar diffusion enabled and with ambipolar
diffusion disabled (i.e., in the ideal MHD limit).

The top grid resolution is $128^3$ with 4 levels of refinement,
resulting in the smallest grid size being $\Delta x = 0.0625\,{\rm
  pc}$.  Grid refinement is based on the requirement that the Jeans
length is resolved with at least 8 zones.  While it has been pointed
out by \citet{2011ApJ...731...62F} that a minimum of 30 zones per
Jeans length is required to resolve turbulence on the Jeans scale,
this is computationally prohibitive given the range of scales in the
problem and the quadratic scaling of the ambipolar diffusion timestep
with grid size.  As the minimum effective grid size is $\Delta x =
0.0625\,{\rm pc}$, the collapse of gas on core-scales is unresolved.
At the maximum refinement level and for $T=10\,{\rm K}$ molecular gas,
the Jeans length is no longer resolved by 8 zones at $3\times
10^3\,{\rm cm^{-3}}$, 4 zones at $1.2\times 10^4\,{\rm cm^{-3}}$, and
1 zone at $2\times 10^5\,{\rm cm^{-3}}$.  This should be kept in mind
when interpreting details of the collapse of individual cores.

\subsection{Heating and Cooling}
\label{Sect:HeatCool}

Heating and cooling are implemented using the Grackle chemistry and
cooling library \citep{2017MNRAS.466.2217S}. The
heating and cooling rates are the same as in
\citet{2015ApJ...811...56W}, although updating to Grackle 2.2 allows
for the use of the mean molecular weight generated by the
PyPDR\footnote{PyPDR is developed by S. Bruderer and is available at
  \url{http://www.mpe.mpg.de/~simonbr/research_pypdr/index.html}} and
CLOUDY \citep{2013RMxAA..49..137F} codes.  In the previous Grackle 2.1
code, the mean molecular weight was determined by an analytic fit as a
function of temperature and was independent of density.  While the
mean molecular weights in the analytic fit agree with the new tabular
values in the cold, dense molecular gas, there is a divergence between
the two at higher temperatures and lower gas densities.  Additionally,
the cooling timestep $\Delta t_{\rm cool}$ is now included in the
total dynamical timestep.  This ensures that gas does not, through
unphysical means, diverge too far from equilibrium temperatures.  For
numerical expediency, however, a minimum cooling timestep $\Delta
t_{\rm cool, min} \sim 600\,{\rm yr}$ is included as the cooling
timestep can become prohibitively small at large densities.  The
cooling timestep takes the form
\begin{equation}
\Delta t_{\rm cool} = \max\left(\frac{0.2 \rho
  e}{|\Gamma-\Lambda|+\epsilon},\Delta t_{\rm cool,min}\right),
\end{equation}
where $\rho$ is the gas density, $e$ is the specific internal energy,
$\Gamma$ is the volumetric heating rate, $\Lambda$ is the volumetric
cooling rate, and the small factor $\epsilon$ is included to avoid
divergence at thermal equilibrium.  This timestep is used in
calculating the dynamical timestep and is independent of the internal
subcycling used by the Grackle cooling routine. These changes result
in a small difference in the gas temperatures at low densities between
results presented in Paper II and those presented here.

\subsection{Ambipolar Diffusion}
\label{Sect:AD}

Ambipolar diffusion is included in \texttt{Enzo}'s Dedner MHD solver
\citep{2002JCoPh.175..645D,2009ApJ...696...96W}\footnote{It is the intent of the authors to make the updated ambipolar diffusion routines available through the \texttt{Enzo} code base available at \url{https://bitbucket.org/enzo/enzo-dev}. } through the
modification of the induction equation to include non-ideal terms:
\begin{equation}
\frac{\partial\vect{B}}{\partial t} - \vect{\nabla}\times\left(\vect{v}\times\vect{B}\right) = -\vect{\nabla}\times\left( D_{\rm AD} \left(\vect{\nabla}\times \vect{B}\right)\times \vect{b}\times\vect{b}\right),
\label{Eqn:Induction}
\end{equation}
where $D_{\rm AD}= c^2\eta_{\rm AD}/4\pi$ is the ambipolar diffusion
constant, $\eta_{\rm AD}$ is the resistivity\footnote{There is no
  consistency within the literature regarding which of the quantities
  $\eta_{\rm AD}$ and $D_{\rm AD}$ is referred to as the
  resistivity.}, and $\vect{b}$ is a unit vector in the direction of
$\vect{B}$.  The terms on the right hand side of equation
\ref{Eqn:Induction} are included as an explicit update in the source
term step.  To ensure
stability, the explicit timestep is limited by the ambipolar diffusion
timestep,
\begin{equation}
\Delta t_{\rm AD} = \frac{\delta\min\left(\Delta x^2,\Delta y^2,\Delta z^2\right)}{D_{\rm AD}},
\end{equation}
where $\delta = 0.1$ is a safety factor included to ensure stability
\citep{1995ApJ...442..726M}.

The inclusion of ambipolar diffusion can introduce time steps much
shorter than the dynamical time step.  \texttt{Enzo} periodically
rebuilds the grid to accommodate the movement of gas.  As the
ambipolar diffusion update does not move gas through the grid, it may
be unnecessary to undergo the expensive process of rebuilding the grid
on the ambipolar diffusion time scale.  To avoid this, the ambipolar
diffusion time step is only included after the time to rebuild the
grid is calculated, thus resulting in the grid being rebuilt on the
ideal MHD dynamical time scale instead of the ambipolar diffusion
timescale.

Additionally, the energy equation is modified to include frictional
heating between the ions and neutrals which, in the strong coupling
limit, takes the form
\begin{equation}
\left(\frac{\partial e\rho}{\partial t}\right)_{\rm AD} =
\frac{4\pi\eta_{\rm AD}}{c} \left| \nabla\times B\right|^2.
\label{Eqn:Energy}
\end{equation}
Previous studies differ on whether such heating is capable of being
comparable to, e.g., the cosmic ray heating rate.
\citet{2000ApJ...540..332P,2012ApJ...755..182P} found that the
ambipolar diffusion heating rate is comparable to the cosmic ray
heating rate for Alfv\'en Mach numbers of order unity; however, this
comparison was made assuming a low cosmic ray ionization rate of
$\zeta_{\rm CR} = 10^{-17}\,\, {\rm s^{-1}}$, much smaller than the
value of $\zeta_{\rm CR} = 10^{-16}\,\,{\rm s^{-1}}$ used here.
\citet{2012ApJ...760...33L} found that turbulent ambipolar diffusion
heating is unlikely to contribute significantly except in the case of
an extremely large, turbulent molecular cloud, and even then, the
heating is likely to be localized to regions affected by shocks.

The ambipolar diffusion heating rate can be estimated from Equation \ref{Eqn:Energy},
\begin{flalign}
\left(\frac{\partial e\rho}{\partial t}\right)_{\rm AD} &=\frac{c^2\eta_{\rm AD}}{16\pi^2}|\nabla \times B|^2   \sim  \frac{c^2\eta_{\rm AD}B^2}{16\pi^2L_B^2} & \\
  &= 1.53\times 10^{-27}\left(\frac{\eta_{\rm AD}}{0.1\, {\rm s^{-1}}}\right)\left(\frac{B}{10\,{\rm \mu G}}\right)^2 & \nonumber \\
  & \times\left(\frac{L_B}{0.0625\, {\rm pc}}\right)^{-2}\,\,{\rm erg\, cm^{-3}\, s^{-1}}  &
\end{flalign}
where $L_B$ is a characteristic scale for magnetic field fluctuations
normalized to the smallest grid scale in our simulations.  While this
estimate does not take into account the character of the turbulence,
it at least shows that heating due to ambipolar diffusion is
potentially influential. 

Standard AD code tests are presented in Appendix \ref{Append:AD}.

\subsection{Calculation of the Resistivity}
\label{Sect:Resist}

The calculation of the resistivity within a cell is done under the
assumption of strong coupling between the ions and neutrals and in the
limit of negligible momentum associated with the ions.  Following
standard derivations (e.g., \citealt{1991pspa.book.....P}), the
conductivities are
\begin{eqnarray}
\sigma_s & = &\frac{n_s e^2 \tau_{s{\rm n}}}{m_s} \\
\sigma_\parallel & = & \sum_s \sigma_s \\
\sigma_\perp & = & \sum_s \frac{\sigma_s}{1+\left(\omega_s\tau_{s{\rm n}}\right)^2} \\
\sigma_{\rm H} & = & -\sum_s \frac{\sigma_s\omega_s\tau_{s{\rm n}}}{1+\left(\omega_s\tau_{s{\rm n}}\right)^2},
\end{eqnarray}
where $\omega_s = qB/mc$ and $\tau_{s{\rm n}}$ is the momentum
exchange timescale for particles of species $s$ with neutral
particles. The associated resistivities are
\begin{eqnarray}
\eta_\parallel & = & \frac{1}{\sigma_\parallel}\\
\eta_\perp & = & \frac{\sigma_\perp}{\sigma_\perp^2 +\sigma_{\rm H}^2}\\
\eta_{\rm H} & = & \frac{\sigma_{\rm H}}{\sigma_\perp^2 +\sigma_{\rm H}^2}.
\end{eqnarray}
As the Ohmic contribution to the resistivity is isotropic, it can be
taken to be equal to $\eta_\parallel$ and then subtracted from the
$\eta_\perp$, yielding the ambipolar resistivity, $\eta_{\rm AD} =
\eta_\perp- \eta_\parallel$.

The resistivity is calculated using the chemical abundances derived
from the chemical model of Paper I.  The momentum exchange timescale
$\tau_{ss'}$ between species $s$ and $s'$ is
\begin{equation}
\tau_{ss'} = \frac{m_s + m_{s'}}{m_{s'}n_{s'}\left<\sigma w\right>_{ss'}}
\end{equation}
where the mass and abundance of species $s$ are given by $m_s$ and
$n_s$, respectively, and $\left<\sigma w\right>_{ss'}$ is the
momentum-exchange coefficient (see Table \ref{TblMomCoeff}). While
other formulations include a correction for the presence of helium,
the model of Paper I contains helium as a separate species and, as
such, collisions between helium and other species are included
separately.

\begin{figure*}
\plotone{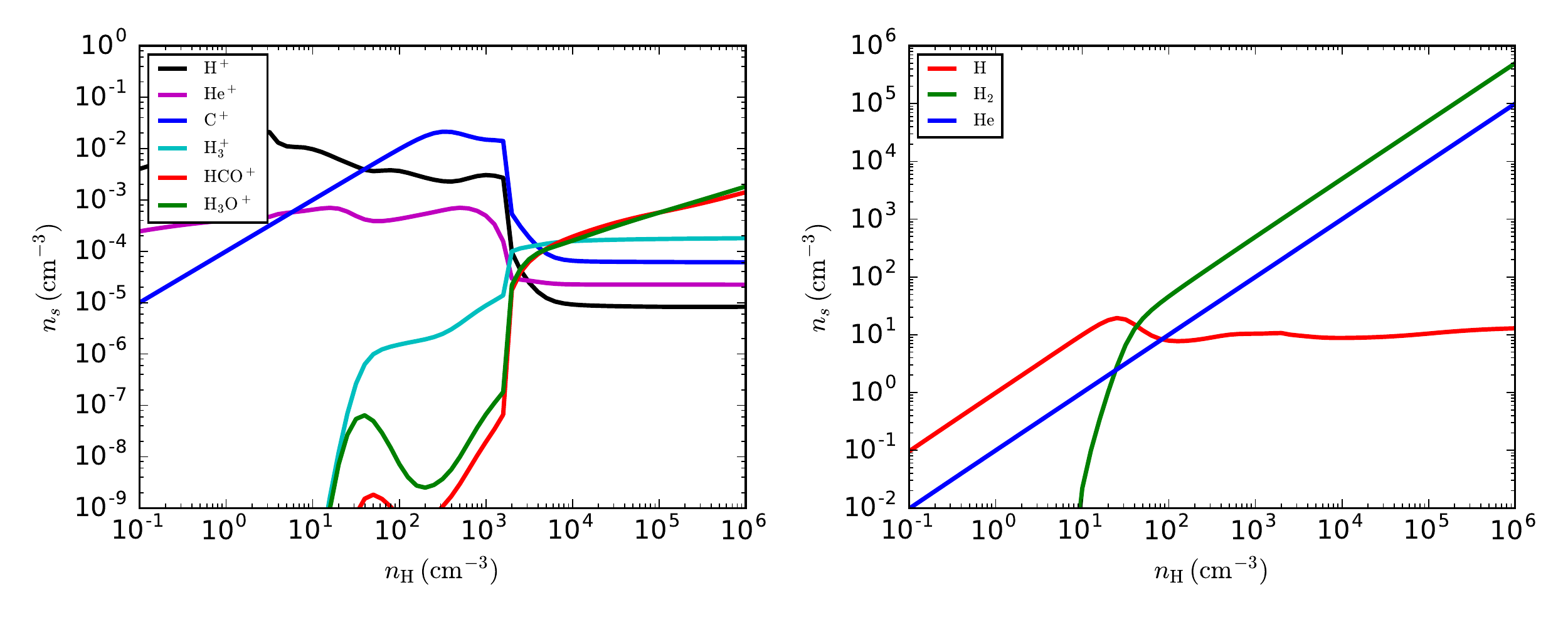}
\caption{
Abundances of the most abundant ions (left) and neutrals (right)
plotted as a function of density, assuming the temperature is that
given by thermal equilibrium. Although still included in the model,
the less abundant ions have been left off for the sake of clarity. }
\label{Fig:ChemAbund}
\end{figure*}

Additionally, neutral species beyond $\rm H_2$, $\rm H$, and $\rm He$
are ignored as their abundances are small and there are uncertainties
in the momentum exchange coefficients.

The neutral gas is assumed to move with a single velocity
$\vect{v}_{\rm n}$, i.e., there is no drift between neutral species.
This allows for the total ion neutral collision timescale $\tau_{s{\rm
    n}}$ to be written as
\begin{equation}
\tau_{s{\rm n}} = \left(\tau_{s{\rm H}}^{-1} + \tau_{s{\rm H_2}}^{-1}
+ \tau_{s{\rm He}}^{-1}\right)^{-1}.
\end{equation}

Table \ref{TblMomCoeff} contains the choices of momentum exchange
coefficients for the various ions within the chemical model, with all
the coefficients being a function of either temperature $T$ or $\theta
= \log(T/K)$.  In cases where a momentum exchange coefficient is not
available for a specific ion, the coefficient for a similar ion is
used.  Specifically, for atomic ions the momentum exchange coefficient
for $\rm C^+$ is used, and for molecular ions the coefficient for $\rm
HCO^+$ is used.

In cases where a momentum exchange coefficient is not available for a
specific neutral species ($n_2$), the Langevin approximation is
employed to relate the missing rate to a known one ($n_1$),
\begin{equation}
\left<\sigma w \right>_{\rm i n_2} = \left[ \frac{\left(m_{\rm i} + m_{n_2}\right)m_{\rm n_1}p_{\rm n_2}}{\left(m_{\rm i} + m_{\rm n_1}\right)m_{\rm n_2}p_{\rm n_1}}\right]^{1/2}\left<\sigma w \right>_{\rm i n_1},
\end{equation}
where the polarizabilities for the neutral species are $p_{\rm He} =
0.207$, $p_{\rm H} = 0.667$, $p_{\rm H_2} = 0.804$
\citep{2008AA...484...17P}.

As the fits in Table \ref{TblMomCoeff} often cover a limited range of
temperatures and with the fits occasionally yielding negative values
at high temperatures, the rate coefficients for $T=1000\,{\rm K}$ are
used when $T > 1000\,{\rm K}$.  This is of negligible impact as the
gas is much cooler in regions where ambipolar diffusion is active.

There are a number of omissions in the calculation of the
resistivity. The chemical model of \citet{2015ApJ...811...56W} does
not contain dust grains as a chemical species, although it does
contain radiative cooling from dust.  For the densities investigated,
it is unlikely that the resistivity will be altered significantly as
the dust grains do not become the dominant charge carrier until
$n_{\rm H_2} \sim 10^7\,{\rm cm^{-3}}$.  The lack of grains also
precludes the possibility of the freeze-out of heavier species such as
$\rm HCO^+$ and $\rm H_3O^+$. Freeze-out of heavier species would
reduce the mean ion mass and thus the coupling of the gas to the
magnetic field. A more thorough discussion of the impact of grains on
the resistivity can be found in \citet{2016MNRAS.460.2050Z} and a
discussion of the effects of non-equilibrium chemistry can be found in
\citet{2012ApJ...753...29T,2012ApJ...754....6T}.  Additionally, the
model also does not have atomic species heavier than oxygen.  Atomic
ions such as $\rm Na^+$ and $\rm Mg^+$ may be important charge
carriers at higher densities.

\begin{figure}[t!]
\plotone{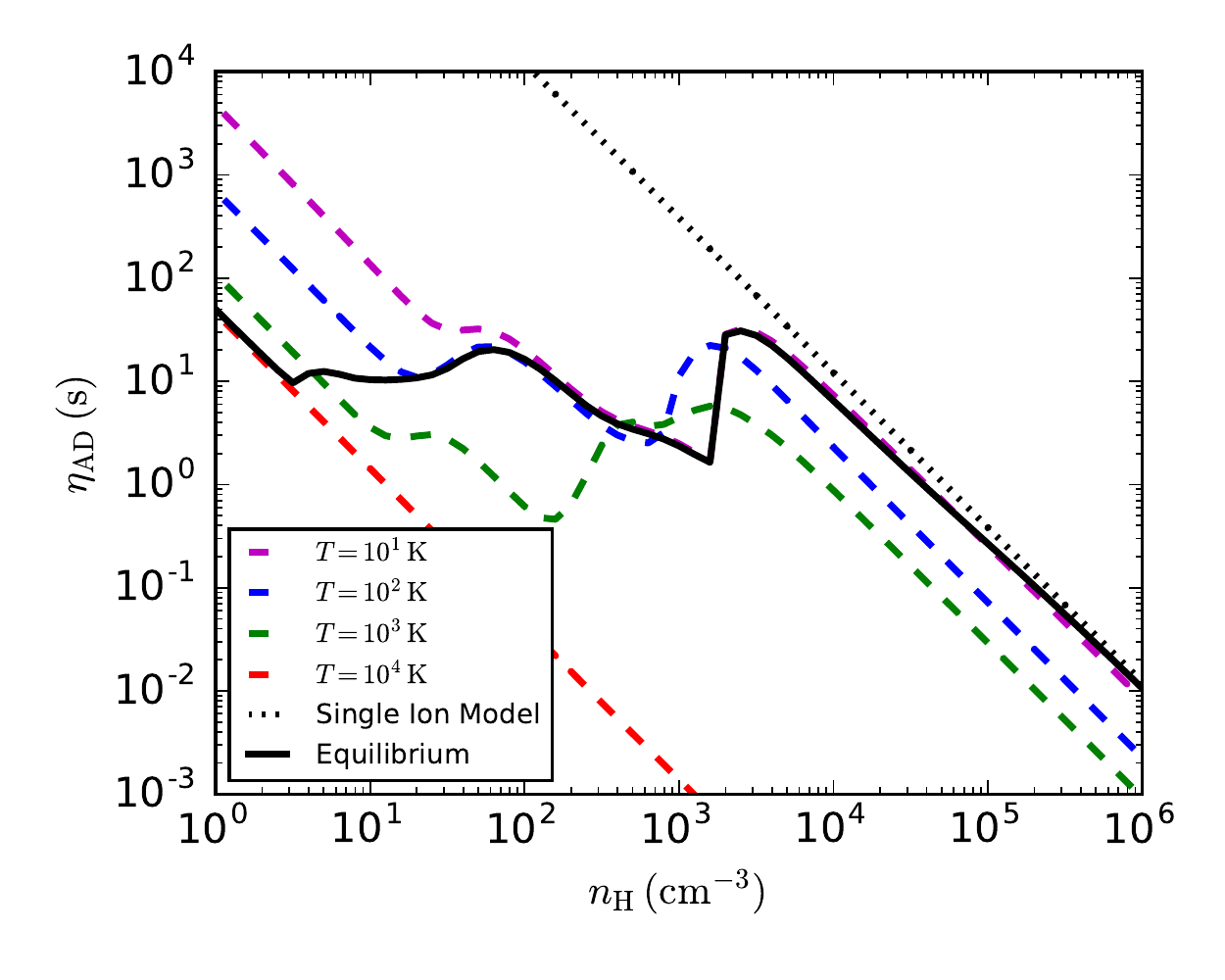}
\caption{
The ambipolar diffusion resistivity $\eta_{\rm AD}$ as a function of
$n_{\rm H}$ for individual temperatures (dashed lines) and along the
equilibrium temperature curve (solid line). The dotted line is
resistivity derived from an isothermal, single-ion model (see text). The magnetic field strength is taken to be $B=10\,{\rm \mu G}$.}
\label{Fig:Resistivity}
\end{figure}

Figure \ref{Fig:ChemAbund} shows the chemical abundances as a function
of $n_{\rm H}$ for equilibrium gas temperatures.  The plot can roughly
be divided into three regimes: the low density gas dominated by
protons and neutral atomic hydrogen, the intermediate range where the
neutral component is now molecular but the dominant ion is $\rm C^+$,
and the final region where the ion density is dominated by $\rm
HCO^+$, $\rm H_3O^+$ and $\rm H_3^+$.

Many works adopt a simplified chemical model wherein cosmic ray
ionization resulting in a representative ion, often $\rm HCO^+$, is
balanced with dissociative recombination \citep{1979ApJ...232..729E}.
Adopting the nomenclature of \citet{2012ApJ...744..124C}, the
resistivity from such a model is given by $\eta_{\rm AD} =
B^2/c^2\alpha \rho_{\rm i} \rho_{\rm n}$ where $\alpha = 3.7\times
10^{13}\,{\rm cm^3\, s^{-1}\, g^{-1}}$ and the the ion number density
is given by $n_i = 10^{-6} \chi_i \left(\rho_{\rm n}/\mu_{\rm
  n}\right)^{1/2}$.  The parameter $\chi_{\rm i} =
10^6\sqrt{\zeta_{\rm CR}/\alpha_{\rm gas}}$ relates the rates of
creation and destruction of the ion species.  The value of $\chi_{\rm
  i} = 3$ used by \citet{2012ApJ...744..124C} is adopted for the
purpose of comparison.

Figure \ref{Fig:Resistivity} shows the computed resistivity for a
number of fixed temperatures in addition to the resistivity along the
equilibrium temperature curve.  At high densities where the gas is
approximately isothermal, there is good agreement with the single ion
model. The models begin to diverge when the primary ions become
atomic, with the models further diverging as the gas temperature
increases at lower densities.

\begin{deluxetable*}{lcll}
\tabletypesize{\footnotesize} 
\tablecolumns{4}
\tablewidth{0pc}
\tablecaption{Momentum Exchange Coefficients \label{TblMomCoeff}}
\tablehead{\colhead{Interaction} & \colhead{$\left<\sigma w\right>_{\rm in}$\,\,$\left(\times 10^{-9}\,\, {\rm cm^3\, s^{-1}}\right)$} & \colhead{Source} & \colhead{Comment}}
\startdata
\sidehead{\em Collisions with $\rm e^-$}
$\rm e^-$-$\rm H$ &  $\sqrt{T}\left(2.841 + 0.093\theta + 0.245\theta^2 - 0.089\theta^3\right)$& \citet{2008AA...484...17P} &  \\
$\rm e^-$-$\rm He$ & $0.428\sqrt{T}$ & \citet{2008AA...484...17P}  & \\
$\rm e^-$-$\rm H_2$ & $\sqrt{T}\left(0.535 + 0.203\theta - 0.163\theta^2 + 0.050\theta^3\right)$ & \citet{2008AA...484...17P} & \\
\sidehead{\em Collisions with ${\rm H^+}$}
$\rm H^+$-$\rm H$ & $0.649 T^{0.375}$ & \citet{2008AA...484...17P}  & \\
$\rm H^+$-$\rm He$ & $1.424 + 7.438\times 10^{-6} T - 6.734\times 10^{-9} T^2$ & \citet{2008AA...484...17P} & \\
$\rm H^+$-$\rm H_2$ & $1.003 + 0.050\theta + 0.136\theta^2 - 0.014\theta^3$ & \citet{2008AA...484...17P} & \\
\sidehead{\em Collisions with $\rm He^+$}
$\rm He^+$-$\rm H$ & $4.71\times 10^{-1}$ & \citet{2004iono.book.....S} & \\
$\rm He^+$-$\rm He$  & $2.0\times8.73\times 10^{-2}(1-0.093\log T)^2$ & \citet{2004iono.book.....S} &\\
\sidehead{\em Collisions with ${\rm H_3^+}$}
$\rm H_3^+$-$\rm H_2$ & $2.693 - 1.238\theta + 0.663\theta^2 - 0.089\theta^3$ & \citet{2008AA...484...17P} & \\
\sidehead{\em Collisions with molecular ions ${\rm M^+}$ other than $\rm H_3^+$}
$\rm M^+$-$\rm H_2$ & $\sqrt{T}\left(1.476 - 1.409\theta  + 0.555\theta^2 -0.0775\theta^3\right)$ & \citet{2008AA...484...17P} & Adopted rate for $\rm HCO^+$\\
\sidehead{\em Collisions with atomic ions ${\rm A^+}$ other than $\rm H^+$, $\rm He^+$	}
$\rm A^+$-$\rm H$ & $1.983 + 0.425\theta - 0.431\theta^2 + 0.114\theta^3$ & \citet{2008AA...484...17P} & Adopted rate for $\rm C^+$\\
\enddata
\tablecomments{The momentum exchange coefficients are functions of both $T$ and $\theta = \log\left(T/K\right)$.}
\end{deluxetable*}

\subsection{Parameter Study}

\begin{deluxetable}{lllll}
\tablecolumns{5}
\tablewidth{0pc}
\tablecaption{Summary of Simulations \label{TblParamStudy}}
\tablehead{
\colhead{Model} & \colhead{Name} &\colhead{$v_{\rm rel}$} & \colhead{B} & \colhead{AD} \\
\colhead{} & \colhead{} &\colhead{$\rm (km\, s^{-1})$} & \colhead{$(\mu G)$} & \colhead{}
}
\startdata
\sidehead{\em Fiducial B-Field Strength}
1 & Ideal Colliding & $10$ & $10$ & No \\
2 & Ideal Non-colliding & $0$ & $10$ & No \\
3 & AD Colliding & $10$ & $10$ & Yes \\
4 & AD Non-colliding & $0$ & $10$ & Yes \\
\sidehead{\em Strong B-Field} 
5 & Ideal Colliding & $10$ & $30$ & No \\
6 & Ideal Non-colliding & $0$ & $30$ & No \\
7 & AD Colliding & $10$ & $30$ & Yes \\
8 & AD Non-colliding & $0$ & $30$ & Yes \\
\enddata
\end{deluxetable}

A parameter study of eight simulations is performed, varying the
magnetic field strength, relative velocity of the clouds (i.e.,
colliding and non-colliding cases), and whether or not ambipolar
diffusion is included.  The run parameters are listed in Table
\ref{TblParamStudy}: the ideal MHD simulations represent a subset of
the parameters investigated in Paper II.  Each simulation is run for
$4\,{\rm Myrs}$, as in Paper II.

\section{Results}
\label{Sect:Results}

Analysis of the runs is performed with the goal of specifically
comparing the ideal and AD simulation results.  Many of the analyses
performed in Paper II resulted in limited differences between ideal
and AD runs and are thus not discussed here.

The visualizations are made in a rotated coordinate system
$\left(x',y',z'\right)$ rotated by $\left(\theta,\phi\right) =
\left(15^\circ, 15^\circ\right)$ relative to the coordinates
$\left(x,y,z\right)$ used in the simulation.  This limits the
contribution to the projected quantities of the planar density
enhancement associated with the colliding ambient medium around the
clouds (note, the ambient gas around each GMC, of ten times smaller
density, is also assumed to be moving with the same velocity as the
GMC). While this enhancement does not contribute significantly to the
dynamics, it can be visible in plots of projected quantities due to
the long path lengths involved.

To better quantify differences between ideal and AD MHD runs, we focus
on density-weighted quantities,
\begin{equation}
\langle X \rangle_\rho = \frac{\int X\rho dV}{ \int \rho dV},
\end{equation}
and their associated variance
\begin{equation}
\sigma^2_X = \langle X^2\rangle_\rho - \langle X\rangle_\rho^2.
\end{equation}
Unless otherwise noted, the integrations take place over the central
$64^3\,{\rm pc^3}$.

\subsection{Cloud Evolution and the Formation of Dense Cores}

\subsubsection{Density Distribution Functions}

\begin{figure*}
\epsscale{1.0}
\plotone{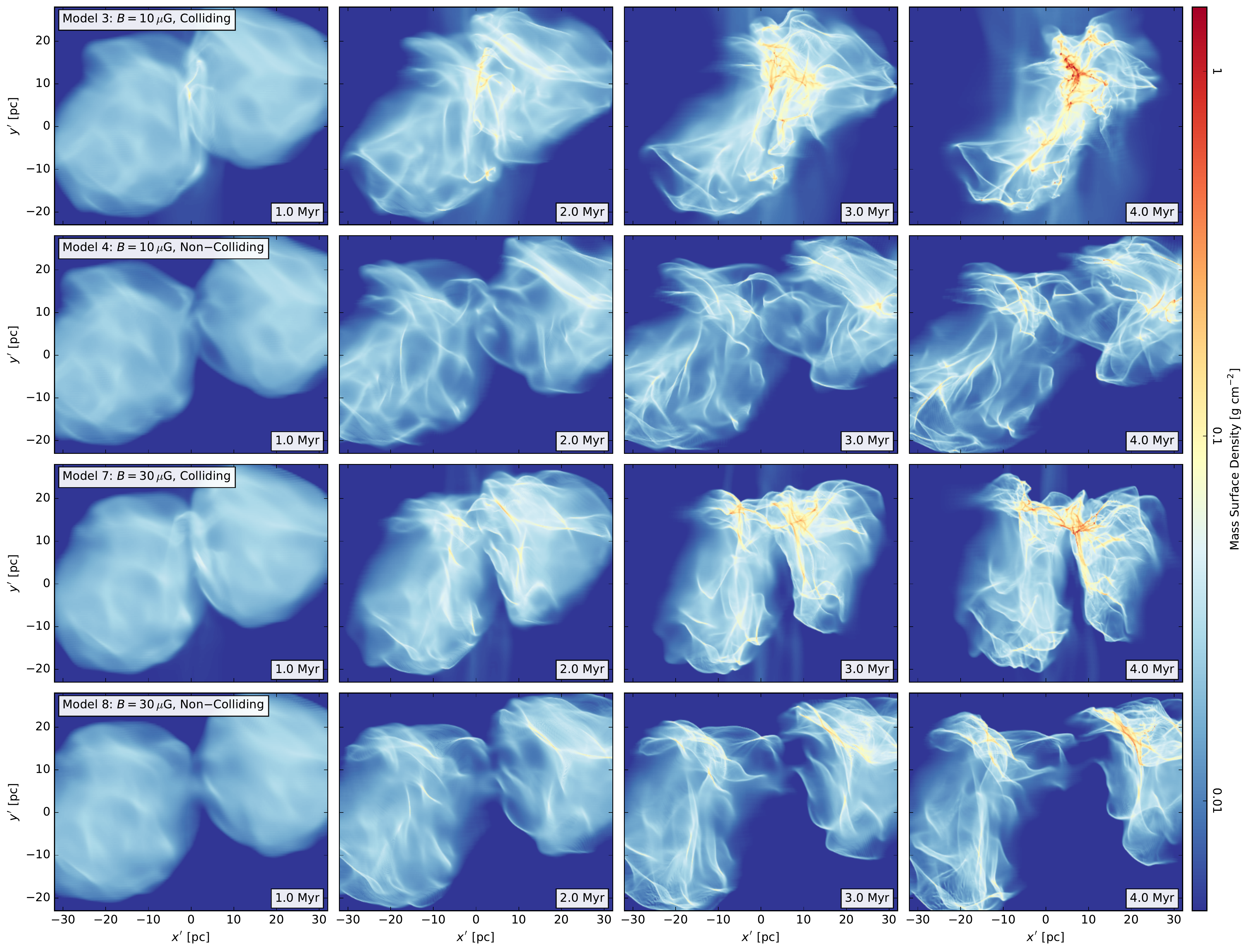}
\caption{
Evolution of mass surface density from 1 to 4~Myr (panels left to
right) for each AD run. First row: $B=10\:{\rm \mu G}$, colliding; 2nd
row: $B=10\:{\rm \mu G}$, non-colliding; 3rd row: $B=30\:{\rm \mu G}$,
colliding; bottom row: $B=30\:{\rm \mu G}$, non-colliding. These
figures show the overall evolution of cloud structures as a result of
internal turbulence and, in the first and third rows, of the 10$\:{\rm
  km\:s}^{-1}$ GMC-GMC collision. Density differences resulting from
AD compared to the ideal MHD case occurring on small scales are presented
in the next figures.}
\label{Fig:EvPlot}
\end{figure*}

Figure \ref{Fig:EvPlot} shows the evolution of the mass surface
density for each of the AD MHD runs, colliding and non-colliding, over
a period of 4~Myr. On these larger scales, cloud evolution progresses
in the same basic manner as in Paper II. In the non-colliding
simulations overdense regions are created by the converging flows and
shocks resulting from the initial internal turbulence that is injected
into the GMCs, as well as a modest degree of collapse caused by the
GMCs being gravitationally unstable. In the colliding simulations,
which Figure~\ref{Fig:EvPlot} shows viewed in a direction
perpendicular to the collision axis, a more substantial amount of
dense gas is created in the regions of the GMCs compressed by the
collision.

For the $10\,{\rm \mu G}$ cases, there are limited morphological
differences between the ideal and AD MHD runs. However, the colliding
$30\,{\rm \mu G}$ ideal and AD MHD runs do exhibit more significant
qualitative differences. In particular, a larger number of dense cores
and filaments are seen to form in the AD simulation. This is
illustrated in Figure \ref{Fig:ColDensCompare}, which zooms in to the
main $\sim$25~pc by $\sim$30~pc concentration of dense gas that is
created by the GMC-GMC collision.  The AD case has $22$ cores
  with densities $n_{\rm H} \ge 10^5\,{\rm cm^{-3}}$, while the ideal
  case only has $15$ cores at $t=4.0\,\,{\rm
    Myrs}$.\footnote{Cores are identified using the the clump
    finding routines in \texttt{yt} \citep{2011ApJS..192....9T} with
    the additional requirement that cores have at least 20 zones.}
The total mass found in the cores is $3880\, {M_{\odot}}$ for the
  AD case compared to $1520\,{M_\odot}$ for the ideal case.  The
larger degree of morphological difference in the $30\,{\rm \mu G}$
ideal and AD runs is expected both due to the increased magnetic
support and the scaling of resistivity with magnetic field strength,
$\eta_{\rm AD} \propto B^2$.

\begin{figure}
\plotone{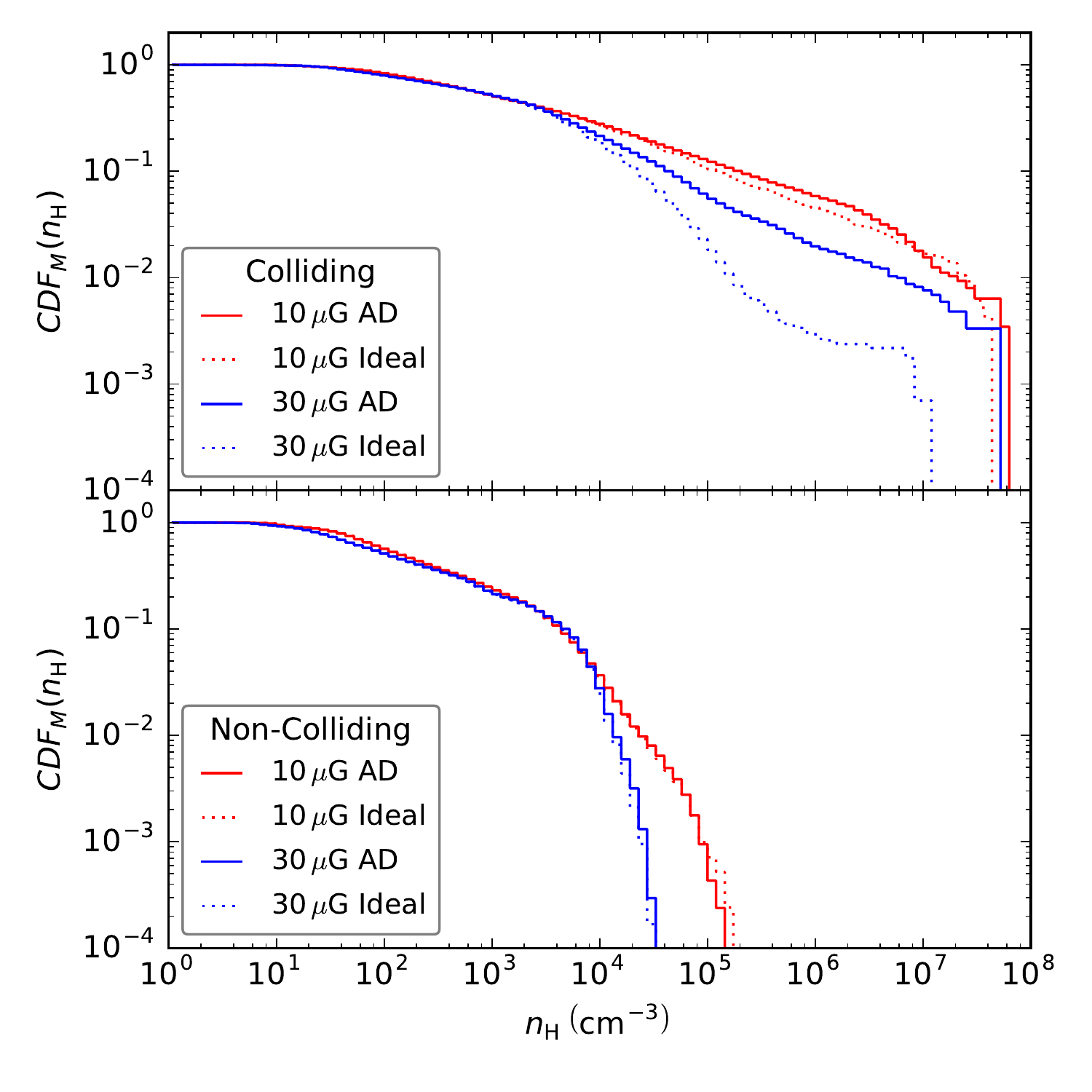}
\caption{Cumulative distribution functions of density for colliding (top
panel) and non-colliding (bottom panel) runs at $t=4\,{\rm Myrs}$.}
\label{Fig:DensHistNoCollide}
\end{figure}

The effects of ambipolar diffusion can also be seen in the cumulative
distribution function (CDF) for the gas density. In general, magnetic
fields act to inhibit collapse, preventing the gas from reaching
higher densities. Thus including ambipolar diffusion should reduce
magnetic support and allow higher densities to be achieved. Figure
\ref{Fig:DensHistNoCollide} 
shows the CDFs for the non-colliding and colliding runs at $t=4$~Myr.

For the non-colliding clouds (Figure \ref{Fig:DensHistNoCollide}, bottom panel), higher densities are created by local
turbulence and gravitational instability set-up as part of the initial
conditions. The GMCs threaded by 10~$\rm \mu G$ $B$-fields achieve
maximum densities of $n_{\rm H}\sim10^5\:{\rm cm^{-3}}$, while those
with 30~$\rm \mu G$ $B$-fields only reach $n_{\rm
  H}\sim3\times10^4\:{\rm cm^{-3}}$. Including AD leads to only modest
changes in the CDFs.

For the colliding GMCs, much higher densities are achieved by the
  end of the simulation runs at 4~Myr, which we note is similar to the
  free-fall time of the initial GMCs, i.e., 4.35~Myr. We consider the
mass fraction above densities of $n_{\rm H}=10^6\:{\rm cm}^{-3}$,
which is a typical density achieved in pre-stellar cores in IRDCs
(e.g., \citealt{2013ApJ...779...96T, 2017ApJ...834..193K,
  2017arXiv170105953K}). In the simulations with $B=10\:{\rm \mu G}$,
in the ideal case this fraction is about 4\%, rising to about 6\% when
AD is implemented. For the $B=30\:{\rm \mu G}$ runs, the effect of AD
is more pronounced. The ideal run has a mass fraction of only 0.2\%
with $n_{\rm H}>10^6\:{\rm cm}^{-3}$, while with AD this increases by
about a factor of 10 to 2\%. These percentages are similar to the star
formation efficiencies per free-fall time inferred from observational
studies of GMCs
\citep[e.g.,][]{1974ApJ...192L.149Z,2007ApJ...654..304K,2016ApJ...833..229L},
however a full quantitative comparison would require simulations
  that explicitly include star formation from the dense gas and
  localized feedback from the newly formed stars.



We note that the differences in the CDFs between these four runs start
to emerge at densities greater than a few~$\times10^{3}\:{\rm
  cm}^{-3}$.
This is likely to be due in part to the local maxima in the ambipolar
resistivity at $n_{\rm H} \sim 3000\,{\rm cm^{-3}}$ (see Figure
\ref{Fig:Resistivity}).

\begin{figure}
\epsscale{1.0}
\plotone{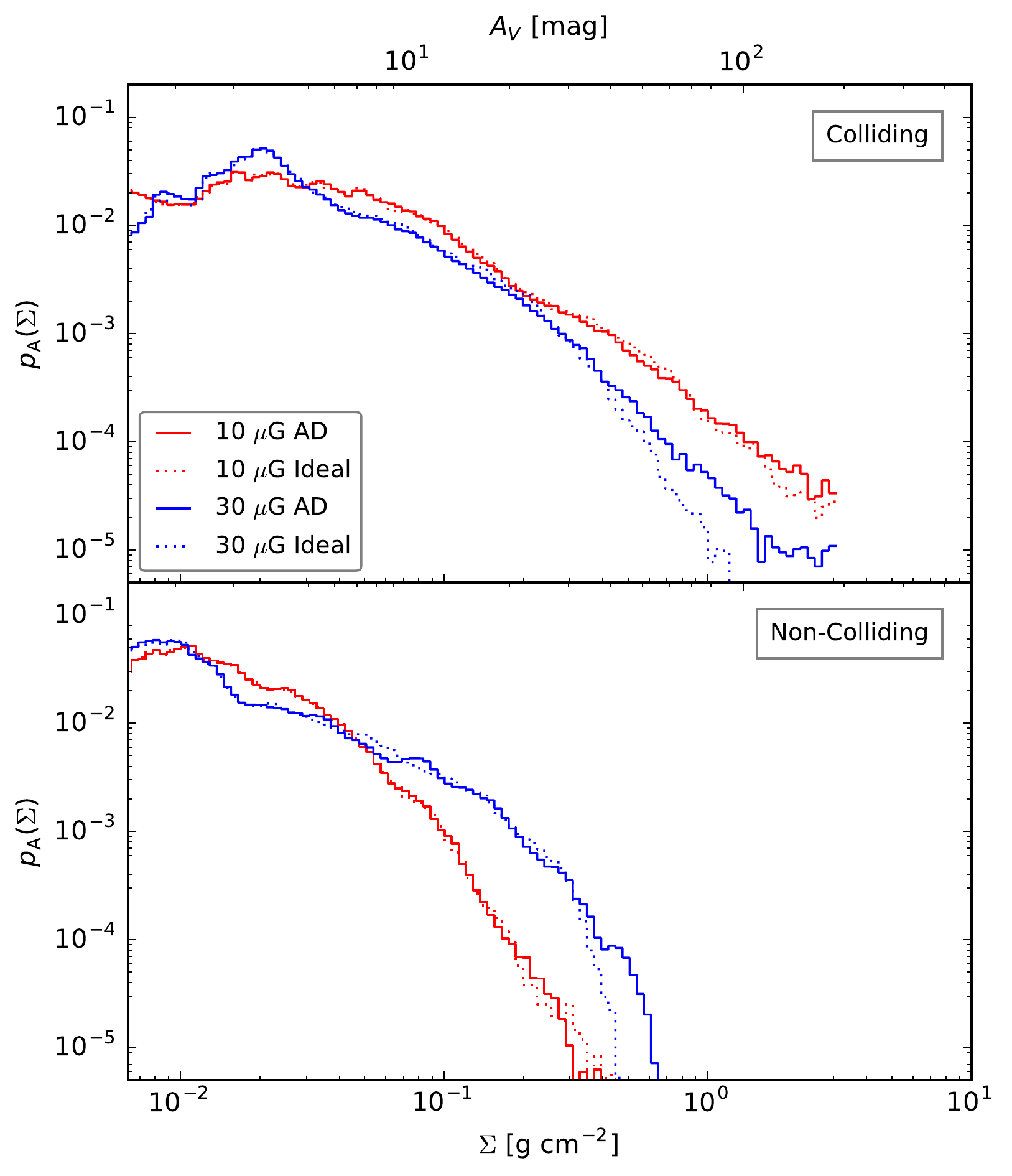}
\caption{The area weighted $\Sigma$-PDFs for the colliding (top panel) and non-colliding (bottom panel) runs at $t=4.0\,{\rm Myrs}$ for the central $(32\, {\rm pc})^2$ regions for each run.}
\label{Fig:SigmaPDF}
\end{figure}

Figure \ref{Fig:SigmaPDF} shows the PDF of mass surface density of the
central $\left(32\,{\rm pc}\right)^2$ regions shown in Figure
\ref{Fig:EvPlot} (and the equivalent regions in the ideal MHD
simulations).  We observe small differences between the ideal and AD
runs for the $10\,{\rm \mu G}$ cases, while in the $30\,{\rm \mu G}$
runs the AD cases consistently have more gas at higher mass surface
densities than in their ideal MHD counterparts, and that the
deviations become prominent at a few$\times 10^{-1}\,{\rm g\,
  cm^{-2}}$.  This is consistent with the results seen in the PDFs of
gas density as well as the qualitative differences seen in Figure
\ref{Fig:ColDensCompare}, which zooms in to part of this region.


In summary, AD can have a significant impact on raising the efficiency
of dense, $n_{\rm H}\gtrsim 10^{6}\:{\rm cm}^{-3}$ gas formation in
the collision of moderately magnetized, but still supercritical,
GMCs. AD would thus also be expected to enhance the star formation
activity that arises in such collisions.

\subsubsection{A Sample Core}

\begin{figure*}
\epsscale{1.1}
\plotone{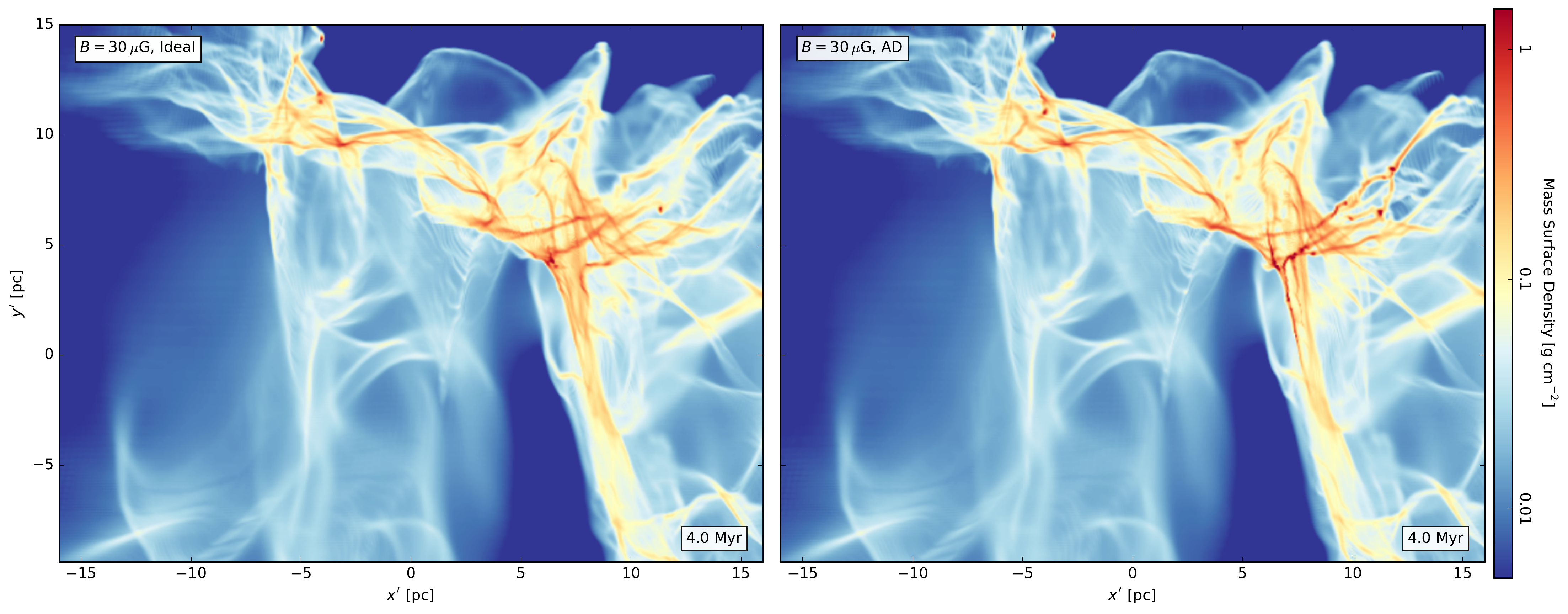}
\caption{
Comparison of the mass surface densities of the $B=30\,{\rm \mu G}$
ideal (Model 5, left) and $B=30\,{\rm \mu G}$ AD (Model 7, right) colliding cloud cases
at $t=4.0\,\,{\rm Myrs}$. The panels show a zoomed-in region centered
on the densest gas structures that are created by the collision. Note
the presence of a larger number of dense cores in the right panel in
the simulation that includes ambipolar diffusion.}
\label{Fig:ColDensCompare}
\end{figure*}

As a case study, we examine a region of Figure \ref{Fig:ColDensCompare} which exhibits core formation in the AD case but no core formation in the ideal case.  This region is not necessarily representative, but is presented to highlight differences between the AD and ideal cases.  

\begin{figure*}
\epsscale{1.0}
\plotone{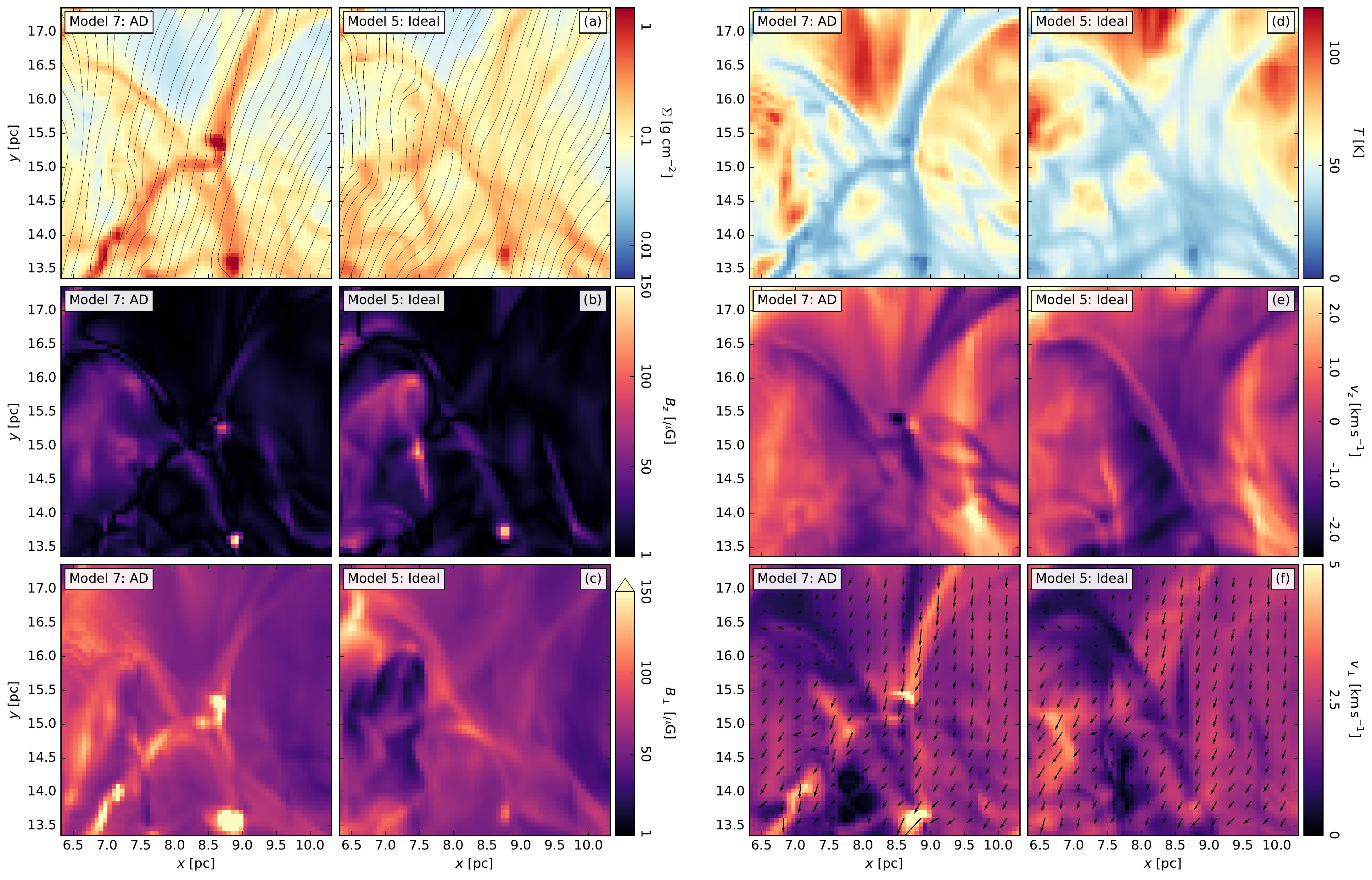}
\caption{
A comparison of a $4\, {\rm pc} \times 4\,{\rm pc}$ region for the AD
case (left panels in each pair) and the ideal case (right panels in
each pair) for the $30\,{\rm \mu G}$ colliding cloud simulations. The
region is selected to highlight collapse and fragmentation in the AD
MHD run that is not found in the ideal MHD run. (a) Mass surface
density. The overplotted streamlines follow the magnetic field
lines. (b) Density-weighted line of sight magnetic field ($B_z$). (c)
Density-weighted plane of sky magnetic field strength. (d) Density-weighted
temperature.  (e) Density-weighted line of sight velocity.  (f)
Density-weighted plane-of-sky velocity.  Vectors indicate the direction
of the flow.}
\label{Fig:CoreZoom}
\end{figure*}

Figure \ref{Fig:CoreZoom} shows a comparison of an example $4\, {\rm
  pc} \times 4\,{\rm pc}$ region for both the ideal and AD $B=30\,{\rm
  \mu G}$ colliding runs.  Panel (a) of this figure shows that the AD
case exhibits an increased number of high mass surface density
($\Sigma \gtrsim 1.0\,{\rm g\, cm^{-2}}$) structures, including a
central core that does not form in the ideal MHD case. This core sits
within a dense, branched filament, which is much more pronounced in
the AD case. This core also reveals itself as a location of stronger
$B$-field strength (panels b and c) and of lower temperature, $T \sim
10\,{\rm K}$ (panel d). Relatively high localized infall velocities
(and velocity gradients), including flows along the filament, are also
present around the core in the AD simulation (panels e and f).

\begin{figure*}
\epsscale{1.0}
\plotone{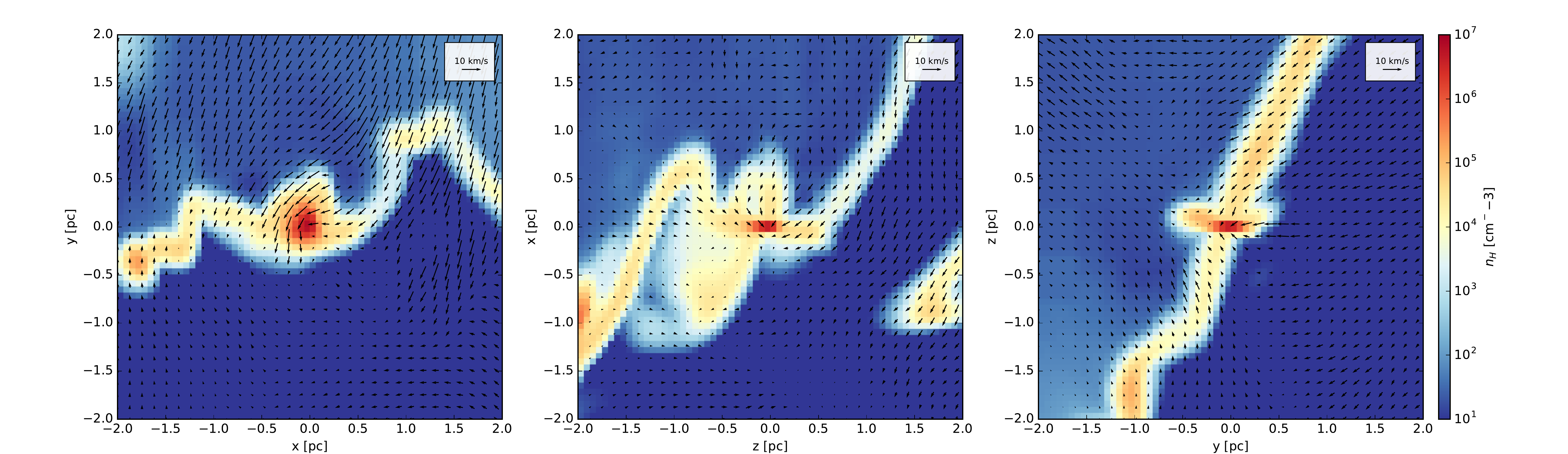}
\caption{
Density slices along each cardinal direction centered on the central
core in the AD simulation (Model 7) in Figure \ref{Fig:CoreZoom}.  Arrows indicate the local
velocity.  The largest projected velocity in each slice is approximately $10\,{\rm km\, s^{-1}}$.}
\label{Fig:CoreSlice}
\end{figure*}

\begin{figure}
\plotone{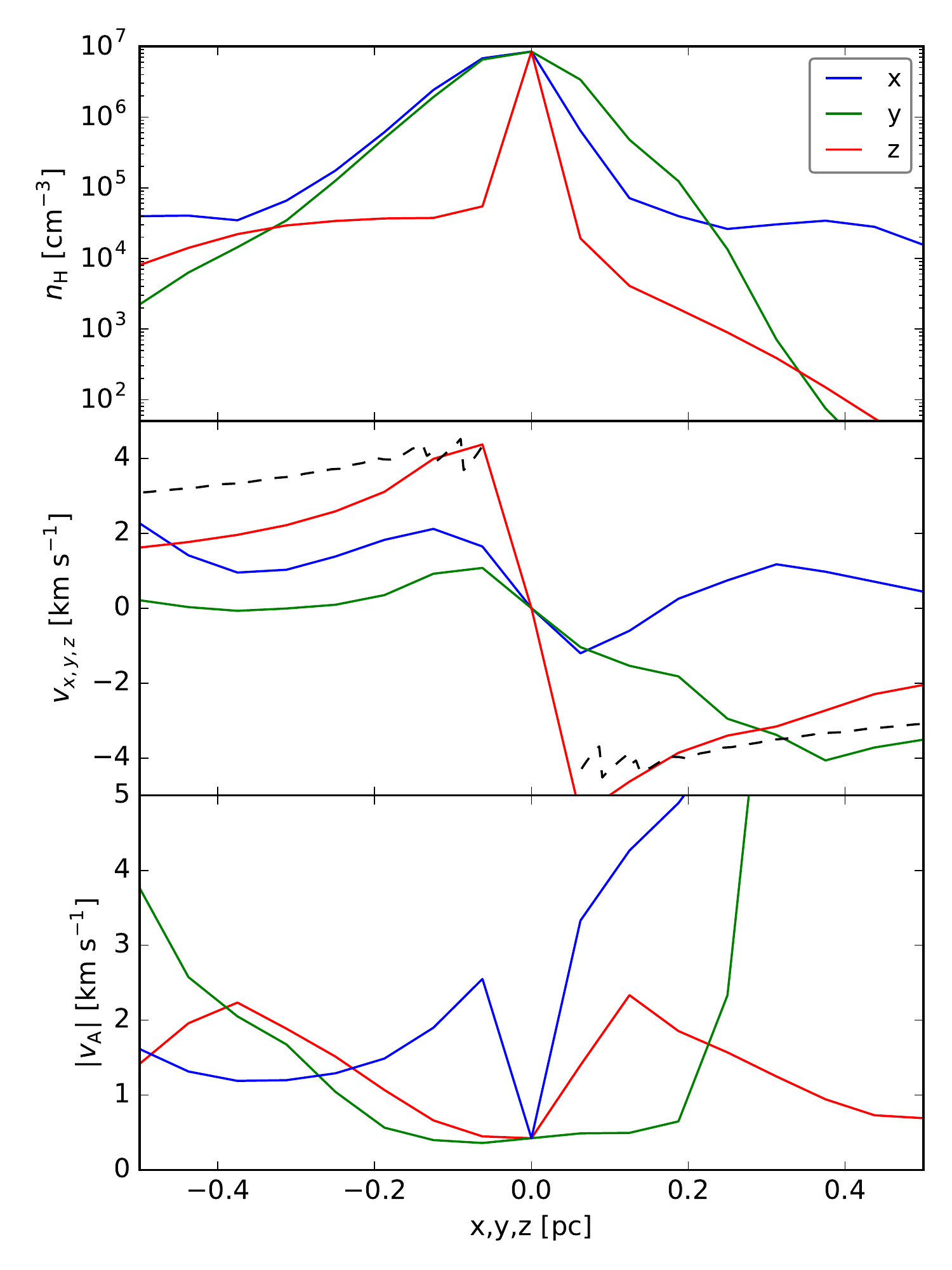}
\caption{
Density (top panel), velocity (middle panel), and local Alfv\'en speed
(bottom panel) profiles along each axis through the center of the core
in Figure \ref{Fig:CoreSlice}.  The coordinates are normalized so that
the density peak is at the origin, and the velocities are in the frame
of the density peak.  In the middle panel, the dashed line shows the
free fall speed $v_{\rm ff} = \sqrt{2 G M_{\rm enc}/R}$.}
\label{Fig:CoreProf}
\end{figure}

Figure \ref{Fig:CoreSlice} shows density slices through the central
core formed in the $30\,{\rm \mu G}$ AD run in the field shown in
Figure \ref{Fig:CoreZoom}. The core exhibits a flattened disk-like
morphology, i.e., being elongated in the $x$-$y$ plane, and appears to
be accreting from surrounding filamentary structures. The velocity
flow patterns generally show infall to the core, but there are also
strong velocity gradients seen in the $x$-$y$ plane indicating a
degree of rotational support.

In Figure~\ref{Fig:CoreProf} we plot density and velocity profiles
along the coordinate axes centered on the core. The relatively
extended density structures in the $x$ and $y$ directions compared to
the $z$ direction are apparent. Infall velocities are supersonic and
trans-Alfv\'enic. However, relative to the free-fall velocity $v_{\rm
  ff} = \sqrt{2 G M_{\rm enc}/R}$ for a sphere of radius $R$ and
associated enclosed mass $M_{\rm enc}$, the velocities along the $x$
and $y$ axes are a factor of several below $v_{\rm ff}$, while along
the $z$ axis, the velocities can be comparable to $v_{\rm ff}$.

\subsection{Magnetic Fields}

\subsubsection{The $B$ vs $n_{\rm H}$ Relation}

\begin{figure*}
\epsscale{1.1}
\plotone{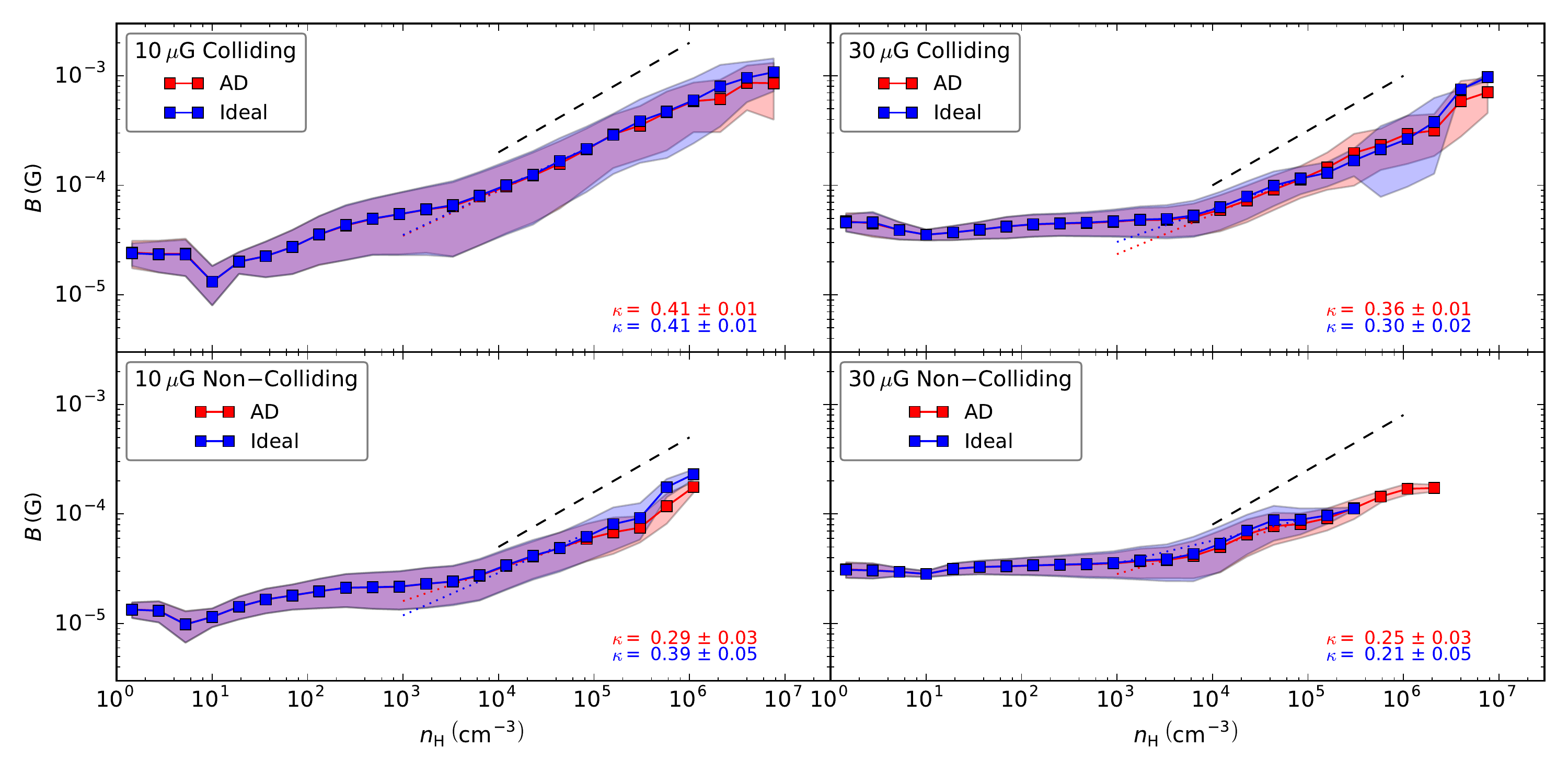}
\caption{
The distribution of magnetic field strengths as a function of gas
density for the four simulation set-ups, each with and without
AD. Solid squares show the density-weighted mean magnetic field for ideal
(blue) and AD (red) MHD runs.  The shaded regions show the $\pm
1\sigma_{\rm B}$ dispersion of magnetic field strengths about this
average. The dashed black line shows the scaling $B \propto n_{\rm
  H}^{1/2}$.  The dotted lines are fits to the respective
density-weighted lines between $n_{\rm H} = 10^3\,{\rm cm^{-3}}$ to
$10^5\,{\rm cm^{-3}}$ with the associated power-law scaling listed in
the lower right corner.}
\label{Fig:Brho}
\end{figure*}

We examine how the magnetic field strength varies with gas density.
Figure \ref{Fig:Brho} shows the density-weighted average magnetic
field strength as a function of $n_{\rm H}$ for the simulation outputs
at 4~Myr. It also shows the dispersion, $\pm \sigma_B$, of the field
strength about this average. The ideal and AD runs show similar
behavior over most of the density range, with deviations confined to
the highest densities, in part because runs with AD are able to
achieve higher densities, i.e., in the non-colliding $B=30\:{\rm \mu
  G}$ case. At low densities, i.e., material external to the initial
GMCs, the field strength is observed to remain approximately
constant. Then in the weaker, $10\:{\rm \mu G}$ initial field case,
the field strength is seen to start increasing at densities of $n_{\rm
  H}\sim100\:{\rm cm}^{-3}$ in the colliding GMCs, but at higher
densities in the non-colliding simulation. In the $30\:{\rm \mu G}$
cases, the gas density above which the $B$-field strength is seen to
start rising is higher still, i.e., at several~$\times10^3\:{\rm
  cm}^{-3}$.

The black dashed lines in Figure \ref{Fig:Brho} show a scaling $B
\propto n_{\rm H}^{\kappa}$ with $\kappa=0.5$, which is the prediction
of some models of clouds with trans-Alfv\'enic turbulence (e.g., Mocz
et al. 2017). Note that clouds with super-Alfv\'enic turbulence are
expected to have a steeper relation with $\kappa\simeq2/3$ (Li et
al. 2015; Mocz et al. 2017). There is expected to be a small decrease
in $\kappa$ due to ambipolar diffusion.  \citet{1993ApJ...415..680F}
found AD causes $\kappa$ to drop from 0.50 to $0.44$-$0.50$
within the context of axisymmetric, isothermal collapse. In our
simulations, the equilibrium temperature decreases from $20\,{\rm K}$
at $n_{\rm H} = 10^3\,{\rm cm^{-3}}$ to $7\,{\rm K}$ at $n_{\rm H} =
10^6\,{\rm cm^{-3}}$.  For polytropes of the form $P \propto
\rho^\gamma$, the expected magnetic field scaling becomes $B \propto
\rho^{\gamma/2}$ which would imply a smaller value of $\kappa$ for gas
that is cooling during its collapse. This effect may explain, at least
in part, why for all our simulation cases, the actual magnetic field
strength scales with an exponent less than $1/2$, i.e., we find
$\kappa\simeq 0.2$ to 0.4 (measured over the range $10^3\:{\rm
  cm}^{-3}<n_{\rm H}<10^5\:{\rm cm}^{-3}$).
  
\subsubsection{The Mass-to-Flux Ratio of Cells}

Paper III developed a star formation subgrid model in which only
cells achieving a particular mass-to-flux ratio $\mu_{\rm cell}$ are
allowed to form star particles.  This model defines the in-cell
mass-to-flux ratio $\mu_{\rm cell}$ as
\begin{equation}
\mu_{\rm cell} = \frac{\Delta x \rho \sqrt{G}}{c_1 B},
\label{Eqn:MTFCell}
\end{equation}
where $c_1$ is the geometry-dependent factor establishing the critical
value, with the adopted value here being $c_1 = 1/\sqrt{63}$ as in
equation \ref{Eqn:MTF}. There are a number of caveats concerning the
quantity $\mu_{\rm cell}$. As defined, $\mu_{\rm cell}$ is inherently
a local, numerical construct and is not related to the stability of a
self-gravitating object. 

The explicit dependence on $\Delta x$ means its value will depend on
the resolution at which it is evaluated.
It does, however, have potential value as input for a subgrid star
formation model, which aims to have sensitivity to the degree of local
magnetic support in the gas.

\begin{figure*}
\epsscale{1.1}
\plotone{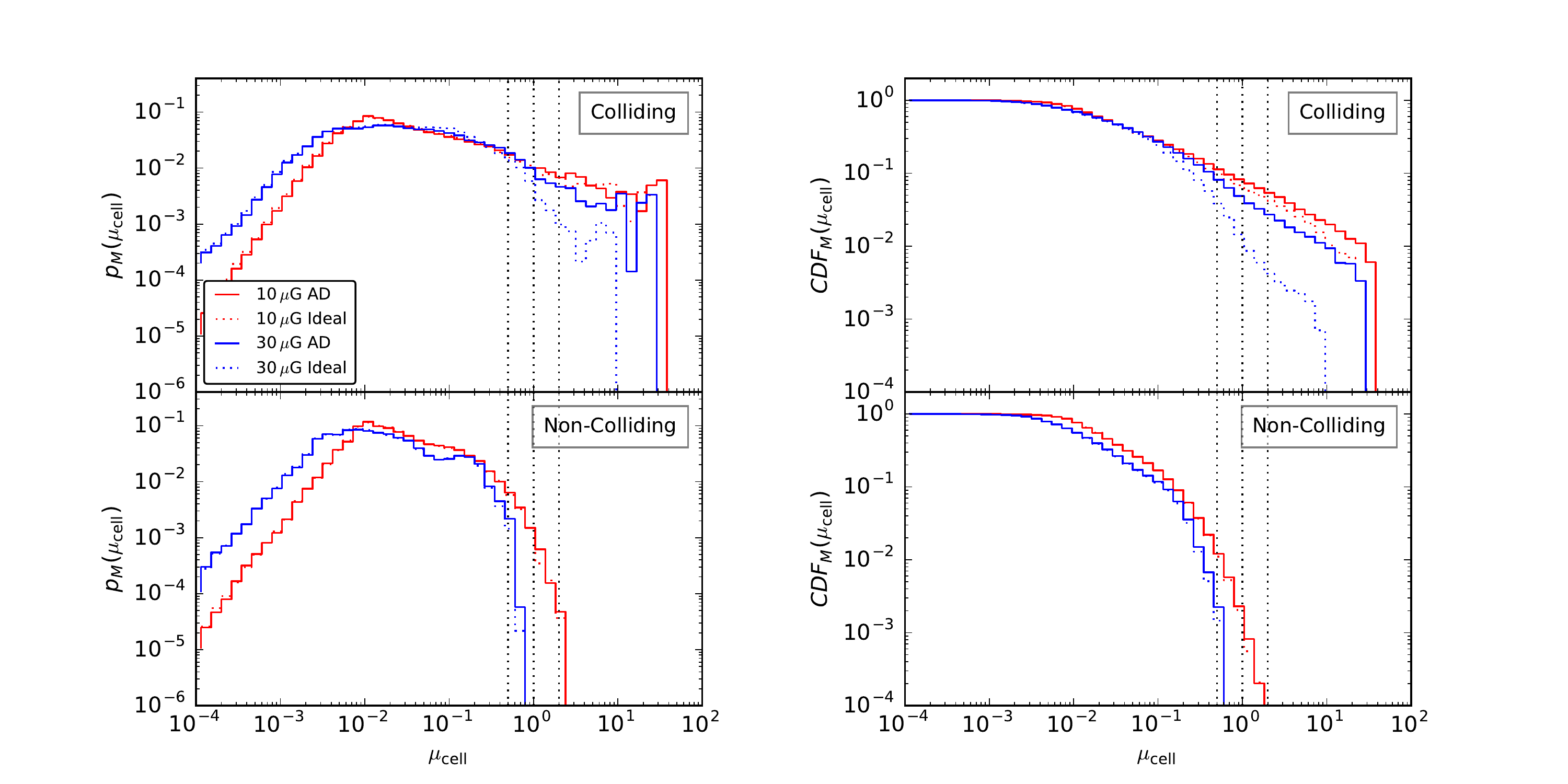}
\caption{
The mass-weighted PDF (left column) and CDF (right column) for the
in-cell mass-to-flux ratios for all colliding (top row) and
non-colliding (bottom row) runs.  The three vertical dotted lines
represent values of $\mu_{\rm cell} = 0.5$, $1.0$, and $2.0$, the
threshold values for star formation used in Paper III.}
\label{Fig:MTFHistCollide}
\end{figure*}

Given that the simulations of Paper III were done in the limit of
ideal MHD, how $\mu_{\rm cell}$ changes in the limit of explicit
non-ideal MHD, i.e., with AD, is of interest.  Figure
\ref{Fig:MTFHistCollide} shows the density-weighted distributions for
$\mu_{\rm cell}$ for the colliding and non-colliding cases.  The
non-colliding cases exhibit relatively minor differences in the
distribution of $\mu_{\rm cell}$ for the ideal and AD cases, as would
be expected given the small changes in volume density distributions.
The $30\,{\rm \mu G}$ colliding case exhibits the greatest differences
between the ideal and AD runs with approximately four times as much
mass in supercritical cells in the AD MHD case compared to the ideal
MHD case. This would lead to correspondingly higher levels of star
formation activity in simulations using such a subgrid
model. Simulations with AD and star formation will be presented in a
future paper in this series.

\subsubsection{Relative Orientations: B vs N}

\begin{figure*}
\epsscale{1.1}
\plotone{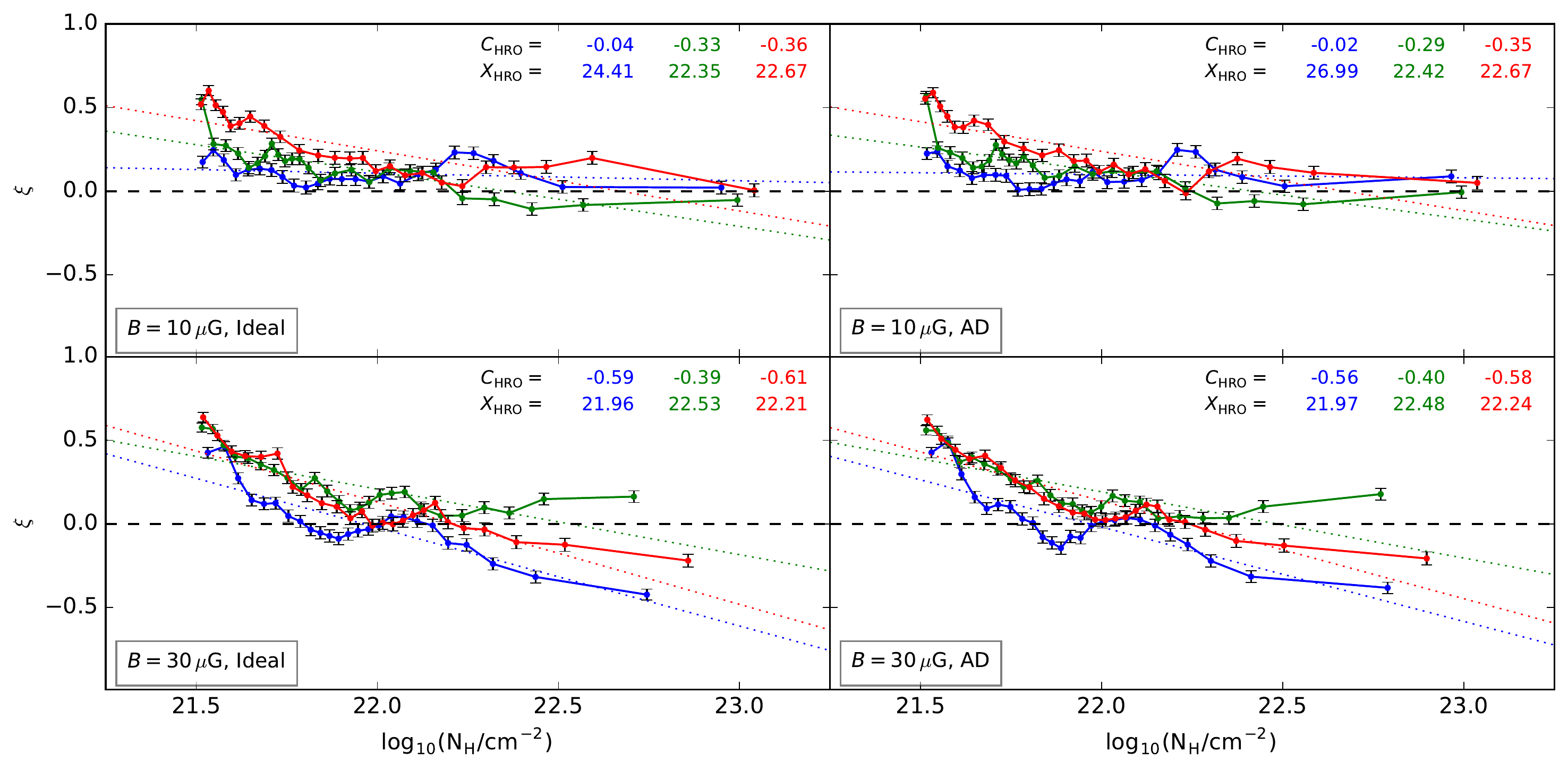}
\caption{The histogram shape parameters $\xi$ vs column density $N_{\rm H}$ for
each of the colliding runs.  The blue, green, and red lines represent
the lines of sight along the $x'$, $y'$, and $z'$ axes, respectively.
The fitting parameters for each line of sight is labelled in its
respective color.}
\label{Fig:HROColliding}
\end{figure*}

As in Paper II, the magnetic field orientations are investigated via
the Histogram of Relative Orientations (HRO,
\citealt{2013ApJS..209...16S}).  The HRO quantifies the relative
orientation between the filamentary structures seen in column density
$N_{\rm H}$ and the plane-of-sky polarization,
\begin{equation}
\phi = \arctan\left(\frac{\vect{\nabla} N_{\rm H}\cdot \vect{p}}{|\vect{\nabla}N_{\rm H}\times \vect{p}|}\right).
\end{equation}
The polarization $\vect{p}$ is a pseudo-vector defined by
\begin{equation}
\vect{p} = \left(p\sin\chi\right)\vect{x} + \left(p\cos\chi\right)\vect{y},
\end{equation}
where $p$ is the polarization fraction and $\chi$ is the polarization
angle.   A discussion of the calculation of $\chi$ can be found in Paper II.
The HRO is then created from the distribution of angles $\phi$.

The convexity of the HRO is defined by the shape parameter,
\begin{equation}
\xi = \frac{A_{\rm c} - A_{\rm e}}{A_{\rm c} + A_{\rm e}},
\label{Eqn:Shape}
\end{equation}
where $A_{\rm c}$ is the area under the central region ($-22.5^\circ <
\phi < 22.5^\circ$) and $A_{\rm e}$ is the area under the extrema
($-90^\circ < \phi < -22.5^\circ$ and $22.5^\circ < \phi < 90^\circ$).
Within this formalism, $\xi > 0$ indicates a concave histogram with
$\vect{p}$ perpendicular to the iso-$N_{\rm H}$ contours while $\xi <
0$ indicates a convex histogram with $\vect{p}$ parallel to the
iso-$N_{\rm H}$ contours.

Figure \ref{Fig:HROColliding} shows $\xi$ as a function of $N_{\rm H}$
for all the colliding runs in the parameter study, along with fits
taking the form
\begin{equation}
\xi\left(N_{\rm H}\right) = C_{\rm HRO}\left[\log\left(N_{\rm H}/{\rm cm^2}\right)-X_{\rm HRO}\right].
\label{Eqn:ShapeFit}
\end{equation}
The same trends found in Paper II for ideal magnetohydrodynamics are
found in the runs in this paper, although the parameters for the fits
are slightly different, which can be attributed to the changes in
numerical methods outlined in previous sections.  Between ideal and AD
runs, there appears to be little difference, except for small
increases in $\xi$ at the highest column densities.

\subsection{Temperature Structure and Resistive Heating}

\begin{figure*}
\plotone{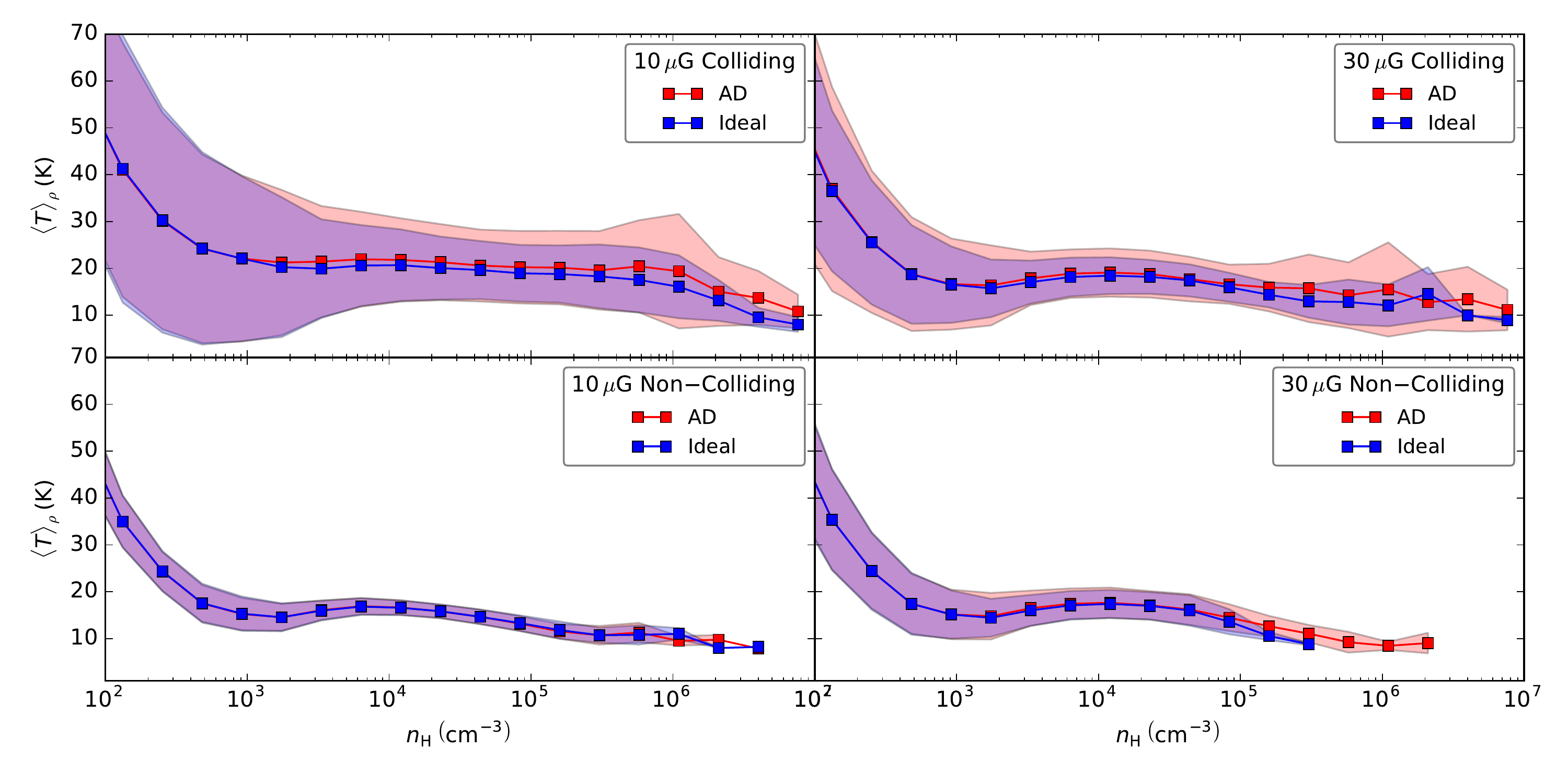}
\caption{
The density-weighted temperature for ideal (blue) and AD (red) MHD runs
(panels, as labelled).  The shaded regions show the $\pm 1\sigma_{\rm
  T}$ dispersion around these average temperatures.}
\label{Fig:Trho}
\end{figure*}

Localized variations of temperature in the ideal and AD runs were
already apparent in Figure~\ref{Fig:CoreZoom}d, but were driven mostly
by the variation in local density structures that arose between the
two simulations. To determine the effect of ambipolar diffusion
heating, we examine the density-weighted temperature $\langle
T\rangle_\rho$ for both the ideal and AD runs.  Figure \ref{Fig:Trho}
shows $\langle T\rangle_\rho$ for each run as well as the $1\sigma$
dispersion about these averages. From comparison with the
  equilibrium temperature versus density relation (Paper I), we see
that the majority of the mass stays close to thermal equilibrium.
However, runs with ambipolar diffusion tend to exhibit a
consistent increase in temperature for $n_{\rm H} \gtrsim 10^3\,{\rm
  cm^{-3}}$, with more significant temperature increases seen in the
colliding runs and/or the more strongly magnetized cases.

Resistive heating has a limited effect on gas at densities below
$10^3\,{\rm cm^{-3}}$.  Compared to their non-colliding counterparts,
colliding clouds exhibit more significant resistive heating, as would
be expected given the magnetic field gradients generated in the
collision and collapse.  Additionally, the stronger field cases
exhibit more significant heating, with temperature increases
approaching $10\%$, as would be expected given the increased
resistivity ($\eta_{\rm AD} \propto B^2$).

As with the differences in the density distribution, the increased
role of resistive heating at $\sim 10^3\,{\rm cm^{-3}}$ is due to the
peak in resistivity caused by the transition from being dominated by
atomic ions to being dominated by molecular ions.

\section{Discussion and Conclusions}
\label{Sect:Conclusions}

In this paper we have developed and implemented ambipolar diffusion in
the \texttt{Enzo} code. We have then investigated the role of
ambipolar diffusion in collisions between GMCs, using resistivities
calculated from the chemical model of Paper I. 

The strongest effects of ambipolar diffusion are seen in the
simulations of colliding GMCs that start with mean $B$-fields of
30~$\rm \mu G$ (i.e., the strong field case), although note that these
GMCs are still magnetically supercritical. The formation of dense,
$\gtrsim10^6\:{\rm cm}^{-3}$ gas is greatly promoted, by factors of
$\sim10$, in the simulation that includes ambipolar diffusion.
One may expect that such differences would also extend to star
formation rates in these GMCs, to be investigated in a future paper in
this series. In the simulations with initial $B$-field strengths of
10~$\rm \mu G$, the GMCs are initially magentically supercritical to a
greater degree and ambipolar diffusion plays a smaller role. 


Resistive heating that arises in the AD simulations also has only a
minor influence on the mass-averaged temperatures, although excursions
to higher temperatures are seen, e.g., in the $1\sigma$ dispersion,
which may have important implications for astrochemical processes,
such as CO freeze-out, normally expected to occur for
$T\lesssim20\:$K.

Since some recent studies of IRDCs have inferred $\sim {\rm m G}$
$B$-field strengths \citep{2015ApJ...799...74P,2016A&A...591A..19P}
and thus trans- or sub-Alfv\'enic turbulence, there may be an
important role for non-ideal MHD processes, such as ambipolar
diffusion, in regulating star formation activity in such
systems. Other non-ideal MHD processes, such as reconnection diffusion
\citep{1999ApJ...517..700L,2011ApJ...743...51E}, may also need to be
considered for a complete understanding of the star formation process
in such systems.



\acknowledgements

Computations described in this work were performed using the
publicly-available \texttt{Enzo} code (http://enzo-project.org), which
is the product of a collaborative effort of many independent
scientists from numerous institutions around the world.  Their
commitment to open science has helped make this work possible. The
authors acknowledge University of Florida Research Computing for
providing computational resources and support that have contributed to
the research results reported in this publication. URL:
http://www.rc.ufl.edu

\bibliographystyle{aasjournal}
\bibliography{colliding-clouds-ad}

\appendix

\section{Tests of the Ambipolar Diffusion Module}

\label{Append:AD}








\subsection{C-Shock Test} 

The case of a C-shock can be solved semi-analytically and has been
used as a test case for implementations of ambipolar diffusion
\citep{1995ApJ...442..726M}. \citet{1995ApJ...442..726M} provide a
semi-analytic solution in the form of a first-order ordinary
differential equation,
\begin{equation}
\left(\frac{1}{\bar{\rho}^2}-\frac{1}{M_{\rm s}^2}\right)L\frac{\partial \bar{\rho}}{\partial x} = \frac{b_y}{M_{\rm A}\left(b_y^2+\cos^2\theta\right)}\left(b_y-\bar{\rho}\left(\frac{b_y-b_{y,0}}{M_{\rm A}^2}\cos^2\theta+\sin\theta\right)\right),
\label{Eqn:CShock1}
\end{equation}
where dimensionless density is $\bar{\rho}=\rho/\rho_0$, and the
dimensionless, perpendicular component of the magnetic field is
$b_y=B_y/B_0$.  $b_{y,0} = \sin\theta$.  The sonic and Alfv\'en Mach
numbers are given by $M_{\rm s} = v_s/c_s$ and $M_{\rm A}=v_s/v_{{\rm
    A},0}$, respectively.  The shock length scale is $L=D/v_{{\rm
    A},0} = B_0/\left(\gamma\rho_i\sqrt{4\pi\rho_0}\right)$.  The
system is closed with the auxiliary equation
\begin{equation}
b_y^2 = b_{y,0}^2 + 2 M_{\rm A}^2\left(\bar{\rho}-1\right)\left(\frac{1}{\bar{\rho}} - \frac{1}{M_{\rm s}^2} \right).
\label{Eqn:CShock2}
\end{equation}
Equation~\ref{Eqn:CShock1} can be integrated to get a final
steady-state solution.

\begin{figure*}
\plotone{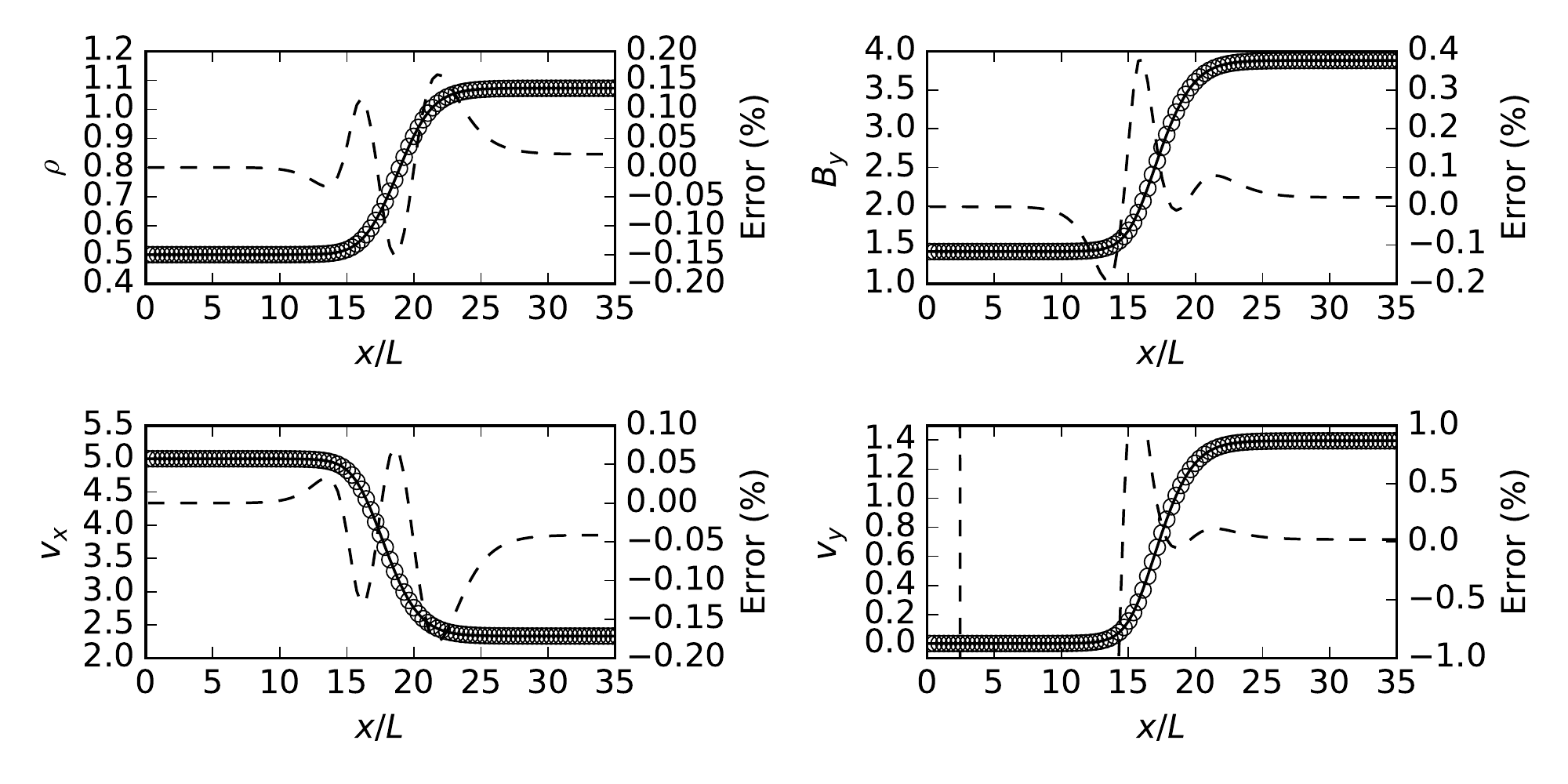}
\caption{The density, perpendicular magnetic field, and velocity components from the C-shock test.  The results of the test are shown as open circles while the solution to Eqn. \ref{Eqn:CShock1} is shown a solid line. The dashed line indicates the relative error. }
\label{Fig:CShockProfile}
\end{figure*}

To generate the C-shock in the simulation, a one-dimensional shock
tube is initialized with a discontinuity at the midpoint and the
upwind and downwind fluid variables given in Table \ref{TblCShock}.
The system is then allowed to evolve and the results are fit to the
semi-analytic solution derived from equations \ref{Eqn:CShock1} and
\ref{Eqn:CShock2}.  The results are shown in Figure
\ref{Fig:CShockProfile} and are in excellent agreement.

\begin{deluxetable}{lcc}
\tablecolumns{3}
\tablewidth{0pc}
\tablecaption{C-Shock Initial Conditions \label{TblCShock}}
\tablehead{\colhead{} & \colhead {Upwind Value} & \colhead{Downwind Value}}
\startdata
$\rho\, ({\rm g\, cm^{-3}})$ & $0.5$ & $1.0727$ \\
$B_x\, ({\rm G})$ & $5.01325$ & $5.01325$\\
$B_y\, ({\rm G})$ & $5.01325$ & $13.7574$ \\
$v_x\, ({\rm cm\,s^{-1}})$ & $5.0$ & $2.3305$ \\
$v_y\, ({\rm cm\,s^{-1}})$ & $0$ & $1.3953$ \\
\enddata
\end{deluxetable}

\subsection{Protostellar Collapse: Isothermal Case}

To set up a protostellar collapse, a cylinder of density
$\rho=1.167\times 10^{-21}\,{\rm g\,cm^{-3}}$ with radius of $0.75\,
{\rm pc}$ and height $1.5\,{\rm pc}$ is placed within a cubic
computational domain of side $2\,{\rm pc}$.  A uniform $20\,{\rm \mu
  G}$ magnetic field is initialized along the cylinder's axis.  This
results in an initial dimensionless mass-to-flux ratio $\mu_0 = 0.55$.

The top grid resolution is $128^3$ with grid refinement done to ensure
that the local Jeans length is resolved with at least 8 zones per
Jeans length.  A maximum of 6 levels of refinement are used resulting
in a top grid size of $0.0325\,{\rm pc}$ and a grid size of
$4.88\times 10^{-4}\,{\rm pc}$ on the finest grid. To facilitate
comparison with other similar non-turbulent collapse models, the
resistivity is taken to be of the form $\eta_{\rm AD} =
\min\left(100,29 B^2/\rho^{3/2}\right)$.  The maximum value placed on
$\eta_{\rm AD}$ both mimics the role of UV in increasing the
ionization and prevents the resistivity from becoming prohibitively
large in low density regions.

\begin{figure*}
\plotone{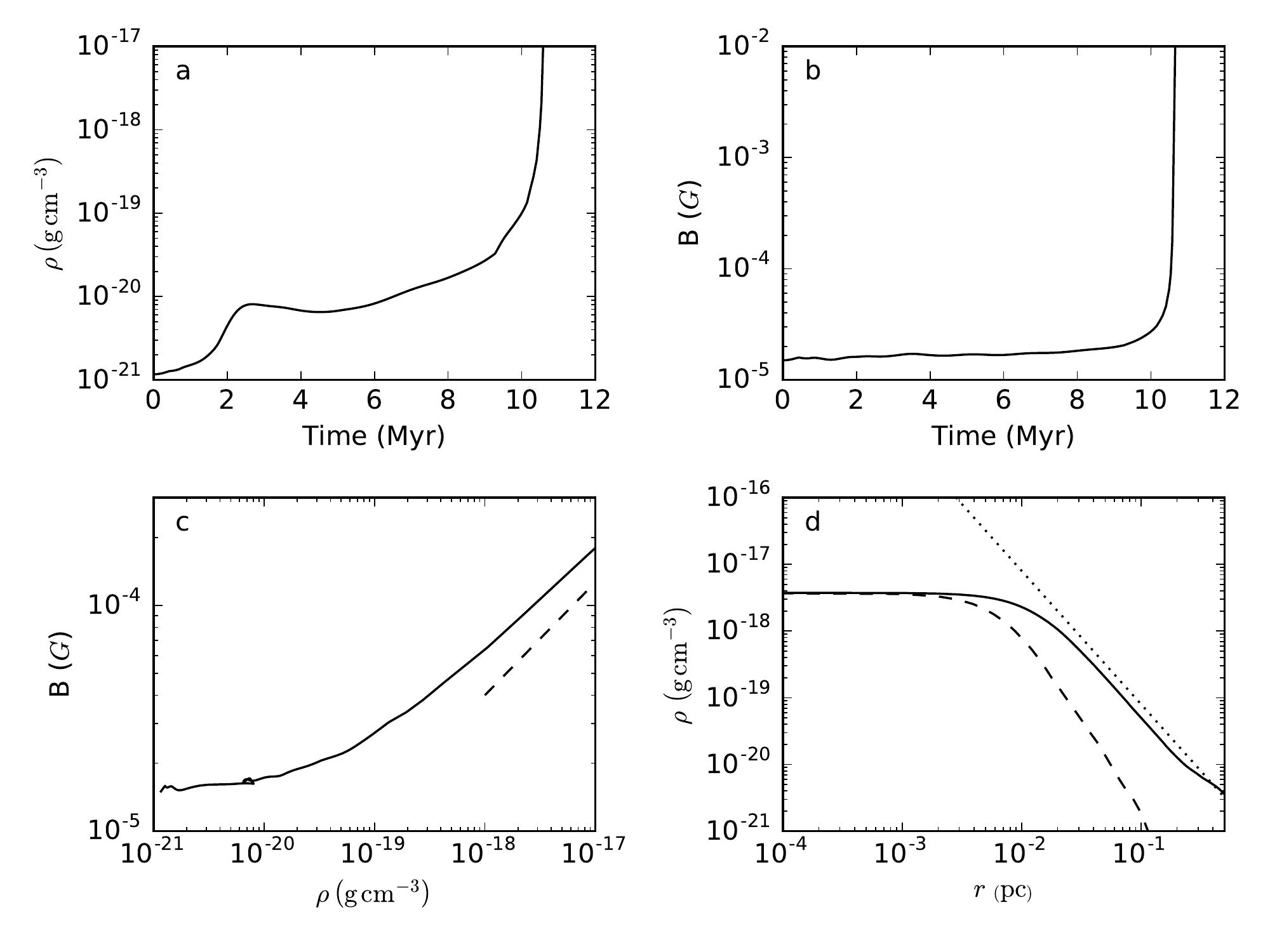}
\caption{
The results the collapse test.  (a) The time evolution of the peak gas
density.  (b) The time evolution of the peak magnetic field strength.
(c) The magnetic field strength as a function of peak gas density
(solid line).  The dashed line shows a scaling of $B \propto
\rho^{1/2}$.  (d) density profiles through the center of the profile
perpendicular to the initial magnetic field direction (solid line) and
parallel to the initial field direction (dashed line).  The dotted
line shows a $\rho \propto r^{-2}$ scaling. }
\label{Fig:Collapse}
\end{figure*}

The results are shown in Figure \ref{Fig:Collapse}.  As has been seen
in similar axisymmetric models, the gas initially collapses along the
magnetic field, resulting in a bounce just after $t=2\,{\rm Myrs}$, as
can be seen in Figure \ref{Fig:Collapse}a.  Via ambipolar diffusion,
gravity draws gas through the field, increasing the central density
while leaving the magnetic field strength roughly unchanged (see
Figure \ref{Fig:Collapse}b).  Once sufficient gas has been accumulated
(around $11\,{\rm Myrs}$), runaway collapse begins with central
density and magnetic field strength increasing dramatically.

Figure \ref{Fig:Collapse}c shows the $B-\rho$ relation for the central
density and magnetic field.  Once dynamic collapse begins, the
magnetic field scales roughly $\rho^{1/2}$, with the scaling being
slightly shallower than the expected $B \propto \rho^{1/2}$ scaling
due to the continued action of ambipolar diffusion during the
collapse.  Figure \ref{Fig:Collapse}d shows the profiles through the
center-of-mass of the cloud at $t=11\,{\rm Myrs}$.  Along directions
perpendicular to the initial magnetic field, the density decreases as
$\rho \propto r^{-2}$, as expected.

\label{lastpage}
\end{document}